\documentclass[A4]{article}
\pdfoutput=1
\usepackage{natbib}
\usepackage[usenames,dvipsnames,svgnames,table]{xcolor}
\usepackage[colorlinks,bookmarksopen,bookmarksnumbered,citecolor=red,urlcolor=red]{hyperref}

\hypersetup{
    pdfauthor={Sylvain Robert},
    pdftitle={Localizing the Ensemble Kalman Particle Filter},
    colorlinks=true,
    citecolor=black,%{blue!10!black},
    linkcolor=black,%
    linkbordercolor=white,
    citebordercolor=white
}

\newcommand\BibTeX{{\rmfamily B\kern-.05em \textsc{i\kern-.025em b}\kern-.08em
T\kern-.1667em\lower.7ex\hbox{E}\kern-.125emX}}

\usepackage{moreverb}

\usepackage{amsmath}
\usepackage{mathtools}
\usepackage{todonotes}
\usepackage{glossaries}
\usepackage{tabularx}
\usepackage[detect-all]{siunitx}
\usepackage[capitalise]{cleveref}
\usepackage{bm} % for bold math
\usepackage{natbib}
\usepackage[T1]{fontenc}

\usepackage{etoolbox}
\robustify\bfseries

\newcommand{\figfactor}{0.7}%{0.47}
\newcommand{\figbig}{0.9}

\begin{document}

%\title{Data assimilation on convective scales: \\ A local ensemble transform Kalman particle filter} % Problem as keywords appear after 65 characters (around transform)!
\title{A local ensemble transform Kalman particle filter for convective scale data assimilation}

\author{Sylvain Robert$^a$,
Daniel Leuenberger$^b$ and Hans R.~K\"unsch$^a$ \\
\small{$^a$Seminar for Statistics, ETH Z\"urich, Switzerland} \\
\small{$^b$Federal Office of Meteorology and Climatology (MeteoSwiss), Z\"urich, Switzerland}}
\date{}

\maketitle

%\corraddr{Seminar f\"ur Statistik, HG G 10.3, ETH Z\"urich, Z\"urich, Switzerland}
%
%\address{\affilnum{a} Seminar f\"ur Statistik, HG G11, ETH Z\"urich, Z\"urich, Switzerland \\ 
%\affilnum{b}Federal Office of Meteorology and Climatology (MeteoSwiss), Z\"urich, Switzerland}

\begin{abstract}
Ensemble data assimilation methods such as the Ensemble Kalman Filter (EnKF) are a key component of probabilistic weather forecasting. They represent the uncertainty in the initial conditions by an ensemble which incorporates information coming from the physical model with the latest observations.
High-resolution numerical weather prediction models ran at operational centers are able to resolve non-linear and non-Gaussian physical phenomena such as convection.
There is therefore a growing need to develop ensemble assimilation algorithms able to deal with non-Gaussianity while staying computationally feasible. 
In the present paper we address some of these needs by proposing a new hybrid algorithm based on the Ensemble Kalman Particle Filter. It is fully formulated in ensemble space and uses a deterministic scheme such that it has the ensemble transform Kalman filter (ETKF) instead of the stochastic EnKF as a limiting case.  A new criterion for choosing the proportion of particle filter and ETKF update is also proposed. 
The new algorithm is implemented in the COSMO framework and numerical experiments in a quasi-operational convective-scale setup are conducted. The results show the feasibility of the new algorithm in practice and indicate a strong potential for such local hybrid methods, in particular for forecasting non-Gaussian variables such as wind and hourly precipitation.
\end{abstract}

%\keywords{Data assimilation, particle filter, ensemble Kalman filter, high-dimensional filtering, convective scale, localization}

\section{Introduction} 

Probabilistic weather forecasts are superior to deterministic ones for a wide range of applications, such as evaluating weather-related risks or managing renewable energy production.
A key element of probabilistic weather forecasting is the use of ensembles methods: instead of running one highly accurate prediction, we can produce  an ensemble of typically 5 to 100 forecasts, which provides information not only on the most probable evolution of the atmosphere, but also on the associated uncertainties. 
Producing such probabilistic forecasts from ensembles is a complex endeavor involving quantification of initial conditions' uncertainties,  representation of model errors, post-processing of ensembles, bias corrections, etc  \citep{gneiting_weather_2005}.
In the present paper, we focus on the production of initial conditions with ensemble data assimilation methods, which combine information coming from the previous weather forecast with the stream of incoming observations. 

Benefiting from increasing computational resources, regional weather forecast models nowadays run with a very high spatial resolution (1 to 5 kilometers), which allows them to resolve small-scales dynamical effects, for example convection \citep{harnisch_initial_2015}. % \citep{gustafsson_convective_2017}. 
On the one hand this is an advantage, as it provides forecasts of high-impact weather events such as heavy storms, but on the other hand it makes the task of data assimilation much harder due to the intrinsic non-linearity of the phenomena resolved at those scales \citep{bauer_quiet_2015}.
Indeed, strong non-linearities 
lead to non-Gaussian uncertainties in the initial conditions, which current methods are poorly equipped to deal with.
Therefore, there is a growing need for computationally efficient ensemble data assimilation algorithms able to handle non-linearities and non-Gaussian distributions.

Ensemble data assimilation methods are sequential algorithms which alternate between two steps. First, during the forecast step, they propagate the ensemble of particles from the previous iteration through the dynamical system, which produces a so-called \emph{background} ensemble. Then, during the update step, or \emph{analysis}, they use the newly available observations to modify the ensemble of particles and produce a so-called \emph{analysis} ensemble.
The various assimilation methods typically differ in the way they implement the analysis.
The current state-of-the-art ensemble methods are based on the ensemble Kalman filter (EnKF)  \citep{evensen_sequential_1994,evensen_data_2009}, 
which conducts the analysis by moving the particles towards the observations in a way that relies implicitly on Gaussian assumptions. 
Particle filters (PFs), on the other hand, directly implement Bayes' formula for the analysis without relying on any Gaussian assumptions \citep{gordon_novel_1993,pitt_filtering_1999,doucet_smc_2001}.
However, this flexibility comes at a cost, and while EnKFs are highly efficient and used in practice, PFs are prohibitively expensive to implement, as they need a very large number of particles to work in high-dimensional systems such as in numerical weather prediction
%, where the dimensionality of the state is of the order of $10^8$ 
(see \cite{snyder_obstacles_2008} for more details on the limits of PFs in high-dimensions). 

Adapting the PF to high-dimensional applications is an active field of research and there have been many propositions of new algorithms, which can be broadly categorized in three different approaches.
The first one is to use variants of the PF with different proposal distributions \citep{pitt_filtering_1999,van_leeuwen_nonlinear_2010,ades_exploration_2013}.
The second is to to create hybrid methods which somehow combine the PF with the EnKF \citep{frei_enkpf_2013,reich_nonparametric_2013}. 
The last approach is to localize the PF, which is difficult but might be the only viable solution for very high-dimensional systems \citep{poterjoy_localized_2016,robert_localization_2016,snyder_performance_2015,rebeschini_can_2015}.
In the present paper we focus on methods which combine the hybrid algorithm approach with localization (see for example \citet{robert_localizing_2017} or \citet{chustagulprom_hybrid_2016}).

Promising results with PFs have been reported on various small- to medium -scale toy models, 
but so far the only application to full-scale weather prediction system that we are aware of is \citet{poterjoy_efficient_2016}.
Here we describe a newly developed localized hybrid algorithm based on the ensemble Kalman particle filter (EnKPF) of \citet{frei_enkpf_2013}. We implemented it in the assimilation framework of the COSMO (Consortium for Small-scale Modeling) model \citep{baldauf_2011}, and we ran successful experiments within the operational data assimilation system of MeteoSwiss.
The implementation of our algorithm was made possible thanks to a collaboration with the Deutscher Wetter Dienst, which is also working on PFs for data assimilation. 

A key development of the new algorithm, called the local ensemble transform kalman particle filter (LETKPF), consisted in formulating the EnKPF in ensemble space, from which we could derive a computationally efficient implementation and a deterministic, or \emph{transform}, analysis scheme.
While other localization methods might be theoretically better \citep{robert_localizing_2017}, we used the  scheme of the local ensemble transform kalman filter (LETKF) \citep{hunt_efficient_2007} for ease of implementation, as it is the assimilation algorithm used by COSMO.
A critical aspect of hybrid methods is to choose the balance between the EnKF and the PF, which is represented by the parameter $\gamma$ in the EnKPF. We proposed and explored a new objective criterion to choose this parameter $\gamma$ adaptively in space and time and compared it to the standard approach.

We conducted numerical experiments with a convective-scale regional model for a period of 12 days in June 2015, with a setup similar to the one used operationally at MeteoSwiss. 
The new algorithm is shown to perform at a similar level to the LETKF, with some noticeable improvements for non-Gaussian variables such as wind and hourly precipitation. These results are very promising for the future of localized hybrid algorithms in challenging real-world applications and we hope that they will spark further interest in our algorithm.

In \cref{sec:background} we review ensemble data assimilation  EnKPF. In \cref{sec:es_algorithm} we derive the new LETKPF algorithm in ensemble space, describe how to compute it efficiently, and discuss how to localize the analysis and choose the parameter $\gamma$ adaptively. In \cref{sec:experiments} we present the numerical experiments and discuss the results of cycled analyses and 24-hour forecasts. \cref{sec:conclusion} concludes with future perspectives.

%%%%%%%%%%%%%%%%%%%%%%%%%%%%%%%%%%%%%%%%%%%%%%%%%%%%%%%%%%%%%%%%%%%%%
\section{Background}
%%%%%%%%%%%%%%%%%%%%%%%%%%%%%%%%%%%%%%%%%%%%%%%%%%%%%%%%%%%%%%%%%%%%%
%%%%%%%%%%%%%%%%%%%%%%%%%%%%%%%%%%%%%%%%%%%%%%%%%%%%%%%%%%%%%%%%%%%%%
\label{sec:background}
The uncertainty about the $q$-dimensional state $x_t$ of a dynamical system based on a stream of observations is best described by probability distributions: The background or forecast distribution $\pi^b_t(x_t)$ is based on observations before time $t$ whereas the analysis distribution $\pi^a_t(x_t)$ includes in addition the current observation $y_t$ according to Bayes' formula: $\pi^a_t(x_t) \propto \pi^b_t(x_t) \cdot \ell_t(x_t|y_t)$
where $\ell_t(x_t|y_t)$ is the likelihood of $x_t$ if $y_t$ has been observed.

Ensemble methods represent these distributions
with finite samples of $k$ particles, $\{ x^{b,i}_t\}$ and 
$\{x^{a,i}_t\}$. These particles are propagated and updated sequentially: 
Propagating $\{x^{a,i}_{t-1}\}$ according to the dynamics of the system produces 
$\{ x^{b,i}_t\}$, updating $\{ x^{b,i}_t\}$
by a sampling version of Bayes' theorem produces $\{ x^{a,i}_t\}$.
Different analysis algorithms vary in the assumptions they make about
$\pi^b_t(x_t)$ and $\ell_t(x_t|y_t)$, and in the sampling version of Bayes'
theorem. 

In the present paper we focus on a single analysis step and thus omit the time
index $t$. We also assume that the observations are linear and Gaussian
with mean $Hx$ and covariance $R$. We next review the EnKPF algorithm in this context and present the EnKF and the PF as special cases.

The EnKPF introduced in \cite{frei_enkpf_2013} 
decomposes the analysis into two stages as $\pi^a(x) \propto \pi^b(x) \cdot \ell(x|y)^{\gamma} \cdot \ell(x|y)^{1-\gamma}$, where $0 \leq \gamma\leq 1$.
%, following the progressive correction idea of \cite{musso2001improving}. 
The core idea of the algorithm is to conduct the first part of the analysis with an EnKF using the dampened likelihood $\ell(x|y)^{\gamma}$, and then to apply a pure PF to the remaining likelihood $\ell(x|y)^{1-\gamma}$.

The first part of the analysis implicitly relies on Gaussianity of the background distribution, but the second part does not make any assumption. 
The EnKPF can thus adapt to some non-Gaussian features of the background distribution without suffering from sample degeneracy like the pure PF.
The parameter $\gamma$ allows one to choose how much of the analysis should be done with the EnKF and how much with the PF, depending on the particular situation at hand.

Using the Gaussian mixture representation of the analysis distribution after the EnKF step, it is possible to derive the final analysis distribution as the following Gaussian mixture:
\begin{equation}
\pi^a_{EnKPF} (x) = \sum_{i=1}^k \alpha^{\gamma,i} \mathcal{N}( \mu^{\gamma,i}, P^{a,\gamma}), \label{eq:pia}
\end{equation}
whose component means $\mu^{\gamma,i}$, mixing weights $\alpha^{\gamma,i}$ and component covariance $P^{a,\gamma}$ are defined as:
\begin{align*}
\mu^{\gamma,i}   &= \nu^{\gamma,i} + K((1-\gamma) Q) (y - H \nu^{\gamma,i} ), \\
P^{a,\gamma}&= \big( I - K((1-\gamma) Q) H \big) Q, \\
\alpha^{\gamma,i}    &\propto \phi\{ y; \ H \nu^{\gamma,i}, HQH' + R/(1-\gamma) \},
% \intertext{where}
\end{align*}
where $\nu^{\gamma,i}$ and $Q$ are intermediary quantities from the EnKF step derived from the background particles $x^{b,i}$ and background covariance matrix $P^b$ as
\begin{align*}
\nu^{\gamma,i}   &= x^{b,i} + K(\gamma P^b) ( y - Hx^{b,i}) \quad \text{and} \\
Q           &= \frac{1}{\gamma} K(\gamma P^b) R K(\gamma P^b)'.
\end{align*}
%are intermediary quantities from the EnKF step.
$K(P)$ denotes the Kalman gain computed using the covariance matrix $P$ and is equal to $P H'(HP H' + R)^{-1}$, while $\phi\{y; \mu, P \}$ denotes the density of a Gaussian distribution with mean $\mu$ and covariance matrix $P$ evaluated at $y$.
More details about the derivation of the EnKPF algorithm can
be found in \citet{frei_enkpf_2013} and \citet{robert_localizing_2017}.

It is convenient for later derivations to rewrite the expression for the $\mu^{\gamma,i}$ components directly from the background ensemble as:
\begin{align}
\mu^{\gamma,i}  &= x^{b,i} + L^{\gamma}(y - Hx^{b,i}), \label{eq:mul} \quad \text{where } \\ 
L^{\gamma} &= K(\gamma P^b) + K((1-\gamma) Q) \Big(I - H K(\gamma P^b) \Big). \label{eq:Lgamma}
\end{align}
$L^{\gamma}$ is the composite Kalman gain resulting from the successive
application of the EnKF and PF. It plays a similar role to the
Kalman gain, but it should be noted that there is no estimate of the
background covariance $P^b$ such that a pure EnKF would have this
gain.

Sampling from \cref{eq:pia} can be done by first sampling the 
indicators $I(i)$ of the mixture components according to 
$P(I(i)=j) = \alpha^{\gamma,j}$ and then adding an artificial noise 
$\epsilon^i \sim \mathcal{N}(0,P^{a,\gamma})$ to $\mu^{\gamma,I(i)}$:
\begin{equation}
\label{eq:xai}
x^{a,i} = \mu^{\gamma,I(i)} + \epsilon^i.
\end{equation}
Instead of sampling with replacement from the set of indices, one can 
generate the indicators $I(i)$ by a
balanced sampling scheme which guarantees that $N^j$, the multiplicity or
number of times a particle $j$ is selected, is less than one unit away from
its expected value, i.e.\ $|N^j - k \alpha^{\gamma,j}| < 1$ 
(for more details on balanced sampling see for example \citet{carpenter_improved_1999}, \citet{crisan_particle_2001} or \citet{kunsch_recursive_2005}).

The EnKF and the PF can be seen as special cases of the
EnKPF. Setting $\gamma$ to 1 we find  
\begin{align*}
\alpha^{1,i} &\propto 1, \\
\mu^{1,i}    &=x^{b,i} + K(P^b) (y - Hx^{b,i}),\\
P^{a,1}   &= K(P^b) R K(P^b)'.
\end{align*}
A balanced sampling scheme, therefore, selects each index exactly once, and thus
we recover the stochastic version of the EnKF. 
At the other end of the spectrum, setting $\gamma$ to 0 we find
\begin{align*}
\alpha^{0,i} &\propto \phi(y;\ Hx^{b,i}, R), \\
\mu^{0,i}    &=x^{b,i},\\
P^{a,0}   &=0.
\end{align*}
The analysis ensemble is thus a resample of the background ensemble with weights proportional to the likelihood, and we recover the PF. For $\gamma>0$, the artificial noise $\epsilon^{i}$ is  not zero and thus no two analysis particles are exactly the same, which is one of the drawbacks of the PF.

%%%%%%%%%%%%%%%%%%%%%%%%%%%%%%%%%%%%%%%%%%%%%%%%%%%%%%%%%%%%%%%%%%%%%
%\section{Ensemble space algorithm}
\section{The local ensemble transform Kalman particle filter}
%%%%%%%%%%%%%%%%%%%%%%%%%%%%%%%%%%%%%%%%%%%%%%%%%%%%%%%%%%%%%%%%%%%%%
\label{sec:es_algorithm}

When the number of particles $k$ is much smaller than the dimension $q$ of the
system, it is desirable that the analysis ensemble belongs to the
ensemble space, i.e.\ the $(k-1)$-dimensional hyperplane in $\mathcal{R}^q$ spanned by the background ensemble. This has advantages both for efficient
implementation and for stability of the assimilation scheme, since
the ensemble space usually contains the main directions of instability.

In the following we represent the background and analysis ensembles as
$q \times k$ matrices $\bm{x}^b$ and $\bm{x}^a$ such that each column
is one ensemble member. The analysis ensemble belongs to the ensemble
space if 
$$x^{a,i} = \sum_{j=1}^k x^{b,j} W_{ji}, \quad \text{ with } \sum_j W_{ji}=1.$$
Equivalently, if and only if the analysis belongs to the ensemble space, it can be expressed as:
\begin{align}
\bm{x}^a = \bar x^b \bm{1}' +  X^b W, \label{eq:XbW}
\end{align}
where $\bm{1}$ denotes the vector of length $k$ with all elements equal to 1, $X^b = \bm{x}^b - \bar x^b \bm{1}'$ the $q \times k$ matrix of deviations from the background mean, and $W$ is a $k \times k$ weight matrix. Because $X^b$ does not have full rank, we do not need to impose the condition $\sum_j W_{ji}=1$.

In order to implement the EnKPF we have to estimate the background
covariance $P^b$. Using the sample covariance matrix 
$$P^b= \frac{1}{k-1} X^b (X^{b})',$$
the resulting analysis is in ensemble space and can be expressed in the form of \cref{eq:XbW}. To prove this and to derive the corresponding $W$ matrix, we first pull out a factor $X^b$ from the matrix $L^{\gamma}$ defined in \cref{eq:Lgamma}
$$L^\gamma = X^b \tilde{L}^\gamma.$$
From \cref{eq:mul} it then follows that the ensemble 
of component means $\mu^{\gamma,i}$ of \cref{eq:pia} is automatically 
in ensemble space:
$$\bm{\mu}^{\gamma} =  \bar x^b \bm{1}' +  X^b W^{\mu}, \quad 
W^{\mu} = I + \tilde L^{\gamma} (y - H\bm{x}^b). \label{eq:Wmu}$$

Resampling of the component means can be described by multiplying $\bm{\mu}^{\gamma}$ from 
the right with the matrix $W^\alpha$, which has exactly one 1 in each column, indicating which particle is resampled, or more precisely
$$
W^{\alpha}_{ij} =
\left\{
\begin{array}{ll}
1 & \mbox{if } I(j)=i, \\
0 & \mbox{otherwise.}
\end{array}
\right.
$$
Therefore, the analysis ensemble from \cref{eq:xai} lies in ensemble space
if the matrix of perturbations $\bm{\epsilon}$ from \cref{eq:xai} can be expressed as
$X^b W^{\epsilon}$:
\begin{equation}
\bm{x}^a = \bar x^b \bm{1}' +  X^b (W^\mu W^\alpha + W^{\epsilon}).\label{eq:xa}
\end{equation}
If we estimate the background covariance by the sample
covariance, we can pull out a factor $X^b$ on both sides of $P^{a,\gamma}$:
$$P^{a,\gamma}= X^b \tilde{P}^{a,\gamma} (X^b)'.$$
Hence in a stochastic version of the filter, we could 
generate $W^{\epsilon}$ as follows
\begin{equation}
W^{\epsilon} = \big( \tilde P^{a,\gamma} \big)^{1/2} E, 
\label{eq:weps}
\end{equation}
where $E$ is a $k \times k$ matrix of centered \emph{i.i.d.}~samples from a
standard normal and $(\cdot)^{1/2}$ is any matrix square-root. Then
$\bm{\epsilon} = X^b W^{\epsilon}$ has exactly mean zero and covariance 
$P^{a,\gamma}$. 

Instead of using a random draw for the added perturbations we would like to use a deterministic scheme for producing $\bm{\epsilon}$. 
The first idea that comes to mind is to redefine $W^{\epsilon}$ in \cref{eq:weps} as the symmetric matrix square-root of $(k-1) \ \tilde P^{a,\gamma}$, because then $\bm{\epsilon}$ has exactly covariance $P^{a,\gamma}$.
However, using such a scheme results in an analysis ensemble with the wrong
covariance, because the $W^{\epsilon}$ generated in this way is strongly
correlated with the $W^{\mu}$ matrix and their effects tend to cancel each
other. For the stochastic version of the filter this problem is not
present because the samples $E$ in \cref{eq:weps} are independent of the
background ensemble. However, for a deterministic filter we need to take
these correlations explicitly into account and match the first and second
moments of the analysis ensemble with their expected values.

The analysis mean, $\bar x^a$, should be equal to  the mean of the resampled component means
\begin{equation}
\bar \mu^{\gamma} =  \sum_{i=1}^k \frac{N^i}{k}  \bm{\mu}^{\gamma,i}. \label{eq:mubar}
\end{equation}
Noticing that $\bar x^b + 1/k \cdot X^b W^{\mu} W^{\alpha} \bm{1} =\bar \mu^{\gamma}$, it is clear that for $\bar x^a$ to equal $\bar \mu^{\gamma}$, $W^{\epsilon} \bm{1}$ must equal $\bm{0}$. In other words, the added perturbations should have mean zero. For the stochastic $W^{\epsilon}$ defined in \cref{eq:weps} this holds because the matrix $E$ is centered such that $E \bm{1} = \bm{0}$. 

The covariance of $\bm{x}^a$ should be equal to the covariance of the resampled component means plus the component covariance:
\begin{equation}
P^{\gamma} =  \sum_{i=1}^k \frac{N^i }{k-1}
(\bm{\mu}^{\gamma,i} - \bar \mu^{\gamma})
(\bm{\mu}^{\gamma,i} - \bar \mu^{\gamma})' + P^{a,\gamma}.\label{eq:Pgamma}
\end{equation}
Computing everything in ensemble space, we can find that for the covariance of $\bm{x}^a$ to equal $P^{\gamma}$, the matrix $W^{\epsilon}$ must satisfy the equation
\begin{align}
A (W^{\epsilon})' + W^{\epsilon} A' + 
W^{\epsilon}(W^{\epsilon})' = (k-1) \tilde P^{a,\gamma}, \label{eq:ricc}
\end{align}
where $A$ is the centered matrix $$A= W^{\mu}W^{\alpha} - \frac{1}{k}W^{\mu} W^{\alpha} \bm{1}\bm{1}'.$$
This is a special form of a continuous algebraic Riccati equation,
or CARE. 
In general, it has infinitely many solutions.
In our experience, requiring $W^\epsilon$ to be symmetric and positive
definite leads to good properties of the analysis. Such a solution 
exists and it can be found efficiently using Newton's method.  Moreover, because of special properties of the matrices involved, it can be shown that this solution of \cref{eq:ricc} guarantees a correct first moment with $W^{\epsilon}\bm{1} = \bm{0}$. Details about the algorithm to solve $W^{\epsilon}$ and the latter property are given in the  \cref{ap:riccati}.
A related algorithm which solves a CARE to obtain an analysis ensemble with correct covariance is described in \citet{de_wiljes_second-order_2016}.

We have thus found a deterministic version of the EnKPF in ensemble space, which we call the ETKPF by analogy with the ETKF, which it is equivalent to when $\gamma=1$.
It should be noted that when $\gamma$ is not equal to 1 the solution found
by the ETKPF is not the same as simply taking the symmetric
square-root of the Gaussian mixture covariance. Indeed, the square-root
scheme is only used as a correction term to the analysis ensemble,
similarly to the random perturbations added in the stochastic
EnKF. In particular the resampling step ensures that interesting
non-Gaussian properties of the analysis distribution are represented in the
ensemble.

%%%%%%%%%%%%%%%%%%%%%%%%%%%%%%%%%%%%%%%%%%%%%%%%%%%%%%%%%%%%%%%%%%%%%
\subsection{Efficient computation}
In principle there are different ways to compute $W$ efficiently, but we chose to follow  the procedure of the ETKF as closely as possible for easy implementation in the COSMO data assimilation framework. 
The starting point is to compute the spectral decomposition of $S$, the weighted covariance matrix of the deviations of the model equivalents $HX^b$ in ensemble space, or more precisely:
\begin{equation}
S = (HX^b)' R^{-1} (HX^b) = U \delta(\bm{\lambda}) U', 
\label{eq:Smatrix}
\end{equation}
where $U$ is the matrix of eigenvectors and $\delta(\bm{\lambda})$ denotes
the diagonal matrix constructed with the vector of eigenvalues
$\bm{\lambda}$. Because $X^b$ is centered, 0 is an eigenvalue of $S$
with eigenvector $\bm{1}$. If the number of observations $d$ is
larger than $k$, $S$ typically has $(k-1)$ non-zero eigenvalues. 

The influence of the observations enters through the following vector:
\begin{equation}
c= (HX^b)' R^{-1} (y - H\bar x^b).
\label{eq:cvector}
\end{equation}

Using Woodbury's formula multiple times and working out the 
algebra, it is
possible to compute $W$ from these elements. For $W^{\mu}$ we obtain the
following expression: 
\begin{align}
W^{\mu} &= U \delta( f^{\mu}(\bm{\lambda})) U' + 
U \delta( f^{\bar \mu}(\bm{\lambda})   )) U'c \bm{1}', \label{eq:wmueff}
\end{align}
where $f^{\mu}$ and $f^{\bar \mu}$ are rational functions and
$f(\bm{\lambda})$ denotes the vector with components $f(\lambda_i)$.
More details about the derivation of this and the following expressions and 
explicit formulas can be found in \cref{ap:deriv}.

The matrix $W^{\alpha}$ does not have to be constructed explicitly, only the weights $\alpha^{\gamma,i}$ and the vector of resampled indices $I$ are needed. Going through the algebra, one can find that the weights are proportional to the following expression:
$$
\exp \Bigg(
-\frac{1}{2} 
\Big(
U \delta( \bm{\lambda} f^{\alpha} (\bm{\lambda}) ) U'
\Big)_{ii} + 
\Big(
U \delta( f^{\alpha} (\bm{\lambda})) U'c
\Big)_i
\Bigg),
$$
where $f^{\alpha}$ is also a  rational function.

Both the stochastic EnKPF and the ETKPF need the ensemble space covariance $\tilde P^{a,\gamma}$ to be computed. Similarly to the calculation of $W^{\mu}$ one can find that 
$$
\tilde P^{a,\gamma} = U \delta( f^{\gamma} (\bm{\lambda})) U',
$$
where $f^{\gamma}$ is another rational function.
For the stochastic EnKPF, the symmetric matrix square root
can thus be computed easily as:
$$\Big( (k-1) \ \tilde P^{a,\gamma} \Big)^{1/2}
= \sqrt{k-1}\cdot U \delta( \sqrt{f^{\gamma} (\bm{\lambda})} ) U'.$$

For the ETKPF one still needs to solve the CARE of \cref{eq:ricc}, which is described in \cref{ap:riccati}, but all its elements can be computed efficiently from the above expressions. 
From these equations we can recover the special cases of the
ETKF and the PF in the limit $\gamma \rightarrow 1$
and $\gamma \rightarrow 0$. Details are given in \cref{ap:deriv}.

%%%%%%%%%%%%%%%%%%%%%%%%%%%%%%%%%%%%%%%%%%%%%%%%%%%%%%%%%%%%%%%%%%%%%
\subsection{Localization}
If the ensemble size is much smaller than the system dimension, all methods described so far perform poorly. The EnKF suffers from spurious long range correlations that result from low rank background covariances. With PFs the  problem is even more pronounced,  as the ensemble collapses if the number of particles does not grow exponentially with the problem size (see \citet{snyder_obstacles_2008} for more detail).
These problems can be overcome by localization, which essentially consists in doing a separate analysis at each site and then \emph{gluing} them together. For the EnKF this is well established, leading to the LETKF and similar methods, However it is not straightforward to use localization for PFs because of discontinuities introduced by resampling different particles at neighboring sites. We now discuss how we address these issues with the EnKPF, which leads to the LETKPF.

The basic idea of localization is to compute different $W$ matrices at every site. For the EnKPF, the $W^{\mu}$ matrices associated with the  component means of the analysis cause no problem as they vary smoothly between adjacent sites, provided that the localization radius is  sufficiently large. In practice, one further enforces smooth transitions by tapering the inverse of the observation covariance matrix $R^{-1}$ as a function of distance. For the EnKPF, however, the biggest issue comes from the resampling matrix $W^{\alpha}$ and the perturbation matrix $W^{\epsilon}$. The problems with the latter are relatively easy to be dealt with, but the ones with the former can only be partially addressed. %Indeed, even if the weights $\alpha^{\gamma,i}$ vary smoothly in space, the resampling step is discrete

In the case of the stochastic EnKPF one simply uses the same noise matrix $E$ to construct $W^{\epsilon}$ in \cref{eq:weps} at every site. Because the covariance matrix $\tilde P^{a,\gamma}$ varies smoothly in space, the $W^{\epsilon}$ matrix constructed in this way does not introduce additional discontinuities. For the ETKPF there is nothing special to do as the algorithm to find $W^{\epsilon}$ is deterministic and its solution varies smoothly between sites.

The main problem comes from the resampling matrix $W^{\alpha}$, which reflects the PF part of the algorithm. 
The weights $\alpha^{\gamma,i}$ vary smoothly in space, but the resampling of particles is discrete in nature and can thus vary abruptly from one location to another.
We now consider three steps to limit the number of discontinuities introduced in this way.

The first step is to reduce the noise added during the choice of the resampled indices vector $I$ from the weights $\alpha^{\gamma,i}$. Clearly, using independent sampling with replacement would be a very poor choice, as even if two adjacent sites had the exact same weights it would result in very different $I$ vectors. The balanced sampling scheme that we use for choosing $I$ is much better as it ensures that the multiplicities of each particle is at most one unit away from their expected value. A simple way to further reduce the added randomness is to use the same random seed at every site. This solution is still suboptimal, but we cannot do better without global communication between sites, which is prohibitive for high-dimensional applications. 

The second step to limit the number of discontinuities is to permute the vector of resampled indices $I$. Indeed, the indexing of particles is arbitrary and can thus be changed without any influence on the local analysis. Unfortunately, finding the optimal permutation of every local $I$ such that the number of discontinuities is minimal is an optimal assignment problem which cannot be solved without using global communication between sites. However, 
putting as many 1 as possible on the diagonal of the $W^{\alpha}$ matrix, and then filling in the remaining cases in a determined order, is simple and reduces discontinuities by a large extent. 

The third step to limit the number of discontinuities is to compute the local analysis on a coarse grid and then to interpolate the  matrix $W$ to a finer grid. This is routinely done with the LETKF in practice, but for different reasons. In the case of the LETKF the main goal is to reduce the computational cost of the analysis, whereas in our case we want to smooth out discontinuities. Let us say we need to match particle $i$ at one coarse grid point with particle $j$ at the next coarse grid point. By interpolating the weights on the finer grid in between, we obtain particles which mix particles $i$ and $j$ progressively, and thus create a smooth transition between both.

It is worthwhile to mention that not all discontinuities are necessarily bad, and it is easy to imagine cases where they are actually positive. In particular if the physical field of interest is not continuous, such as a cloud field, it makes sense to match together different particles at different sites. The problems arise when the estimated derivatives in the propagation step become large, which can result in gravity waves or other spurious dynamical effects. The extent to which such harmful discontinuities are avoided with our algorithms needs to be studied in practice.

%%%%%%%%%%%%%%%%%%%%%%%%%%%%%%%%%%%%%%%%%%%%%%%%%%%%%%%%%%%%%%%%%%%%%
\subsection{Adaptive choice of $\gamma$}
\label{sec:adaptive_gamma}
The parameter $\gamma$ determines the proportion of the analysis done with the EnKF and with the PF. There is no reason to fix it a priori and we would like a criterion to select its value adaptively.
\citet{frei_enkpf_2013} proposed to choose the smallest $\gamma$ such that the equivalent sample size (ESS) \citep{liu_metropolized_1996}, computed from the mixture proportions as $1/\sum (\alpha^{\gamma,i})^2$, is within a given bound, for example no less than 50\% of the original ensemble size. This idea is reasonable and particularly cheap to implement, but the problem of choosing $\gamma$ is transfered to the problem of choosing the desired reduction in equivalent sample size, and it does not provide us with a clear criterion for the latter. In \cref{sec:experiments} we use this criterion with a targeted ESS of 50\% as a reference to which we compare the alternative solution proposed below.

Another approach that seems attractive at first sight is to make $\gamma$ a function of the ``non-Gaussianity'' of the distribution. The motivation is that if the background ensemble is truly Gaussian, one should choose $\gamma=1$ and recover the EnKF, while the more non-Gaussian the distribution, the more PF should be used. However, there are at least two reasons why this idea is not applicable in practice. First, the concept of non-Gaussianity is not well defined, as there are infinitely many ways for a distribution to be non-Gaussian, especially in higher dimensions; but even with a measure of non-Gaussianity, one would still have to map its value to a choice of $\gamma$ between 0 and 1, for which we would still have no guidance. Second, there are cases where the background distribution is clearly non-Gaussian but it might be preferable to choose a $\gamma$ close to 1. Indeed, if the observation $y$ is situated outside of the convex hull formed by the ensemble, the weights $\alpha^{\gamma,i}$ will be very skewed and thus lead to sample depletion. In such a case we would be better off choosing a large $\gamma$ even if the background is non-Gaussian.

To address the various points above we propose to base the choice of
$\gamma$ on the mean squared error (MSE) of the predictive mean of $y$. 
From \cref{eq:pia} it follows that the predictive distribution of $y$ is the following mixture:
\begin{equation}
\pi^a(y) = \sum_{i=1}^k \alpha^{\gamma,i} \mathcal{N}( H \mu^{\gamma,i}, HP^{a,\gamma}H' + R). \label{eq:piay}
\end{equation}
In order to take into account the error coming from the resampling step, 
we condition on the multiplicities $N^i$ (the number of times component $i$ is resampled), and consider the following predictive distribution:
\begin{equation}
\pi^a(y | \{N^i\}) = \sum_{i=1}^k \frac{N^i}{k} \mathcal{N}( H \mu^{\gamma,i}, HP^{a,\gamma}H' + R), \label{eq:piayI}
\end{equation}
whose mean is $H \bar \mu^{\gamma}$ given in \cref{eq:mubar}.

We then choose $\gamma$ such that the MSE of the predictive mean, $H \bar \mu^{\gamma}$, is minimal. 
Because the observations do not all have the same variance, it is necessary to scale the MSE with $R^{-1}$, or more precisely:
\begin{equation}
\text{MSE}(H\bar \mu^{\gamma}, y) = (y - H \bar \mu^{\gamma})' R^{-1} (y - H\bar \mu^{\gamma}), \label{eq:mse}
\end{equation}
where the predictive mean $H \bar \mu^{\gamma}$ depends on
$\gamma$. Writing $\bar \mu^{\gamma}$ as $\bar x^b + X^b m^\gamma$, where 
$m^\gamma$ is
the weight vector defined by $\frac{1}{k} W^{\mu} W^{\alpha} \bm{1}$, the
MSE above can be written as:
\begin{equation}
\text{MSE}_{\gamma}(H\bar \mu^{\gamma}, y)  = 
\text{MSE}(H\bar x^b,y) + (m^\gamma)' S m^\gamma - 2 (m^\gamma)'c
\label{eq:mse_es}
\end{equation}
where $S$ and $c$ are defined in \cref{eq:Smatrix} and \cref{eq:cvector}. 
Since the first term is independent of $\gamma$, 
we can choose $\gamma$ adaptively by minimizing 
$(m^\gamma)' S m^\gamma - 2 (m^\gamma)'c$, for example with a grid search.

The scheme for choosing $\gamma$ proposed above is objective and does not
need any additional tuning parameter. On the other hand, it might lead to
over-fitting as it uses the observations $y$ twice: once for computing the
analysis given $\gamma$ and once for computing the MSE.
Practical experiments are needed to evaluate if this is a non-negligible effect. One potential remedy to mitigate the problem is to use the jackknife, a bias reduction technique, to estimate the expected MSE. 
A more radically different approach would be to use a cross-validation scheme with surrogate data created from the background ensemble. The latter approach is attractive from a theoretical point of view, but it is computationally expensive and implicitly relies on the assumption that the ensemble and the truth are exchangeable, which might be violated in case of systematic model biases.

Instead of the MSE of the analysis mean, we could also use the energy score (ES),  a strictly proper multivariate generalization of the continuous ranked probability score (CRPS) \citep{gneiting_strictly_2007}. We developed an algorithm to approximate the ES in ensemble space but the resulting choice of $\gamma$ was not significantly different from using the MSE criterion above, and we thus prefer the latter method for its simplicity. 
Optimal selection of the parameter $\gamma$ depends on many different variables such as the number of observations compared to $k$, the distribution of the background and the assimilation strength, and should be the object of further research.

%%%%%%%%%%%%%%%%%%%%%%%%%%%%%%%%%%%%%%%%%%%%%%%%%%%%%%%%%%%%%%%%%%%%%
\section{Numerical experiments}
%%%%%%%%%%%%%%%%%%%%%%%%%%%%%%%%%%%%%%%%%%%%%%%%%%%%%%%%%%%%%%%%%%%%%
\label{sec:experiments}
The new algorithms described above were implemented and tested in practice on a quasi-operational setup at MeteoSwiss. We first describe the experimental setup in \cref{sec:setup} and then discuss the main results in \cref{sec:results}.

%%%%%%%%%%%%%%%%%%%%%%%%%%%%%%%%%%%%%%%%%%%%%%%%%%%%%%%%%%%%%%%%%%%%%
\subsection{Experimental setup}
\label{sec:setup}
In this section we briefly introduce the KENDA system used at MeteoSwiss before describing the test period and the experiments.

\subsubsection{The KENDA system}
\label{sec:kenda_system}
The numerical experiments in this study were carried out using the KENDA (Kilometer-Scale Ensemble Data Assimilation) system as described in \citet{schraff_2016}. It is based on the COSMO model \citep{baldauf_2011} with a setup similar to the operational implementation at MeteoSwiss.

The COSMO model is a convective-scale, non-hydrostatic NWP model developed within the COSMO consortium (http://cosmo-model.org) and operated at many national weather services worldwide. The atmospheric prognostic variables are the three-dimensional wind, temperature, pressure, turbulent kinetic energy and specific contents of water vapor, cloud water, cloud ice, rain, snow and graupel. The equations for the dynamic variables are solved using a Runge-Kutta time-splitting scheme. Deep convection is explicitly computed, whereas shallow convection is parametrized. A one-moment Lin-type cloud microphysics scheme is responsible for the conversions among all cloud and hydrometeor types. The turbulence parameterization is based on the prognostic Turbulent Kinetic Energy (TKE) equation and radiative effects are parametrized using a $\delta$-two-stream scheme. A multi-layer soil model provides the lower boundary condition at the ground. For more details of the COSMO model we refer to \citet{baldauf_2011}. 

\begin{figure}
    \centering
    \includegraphics[width=\figfactor\textwidth]{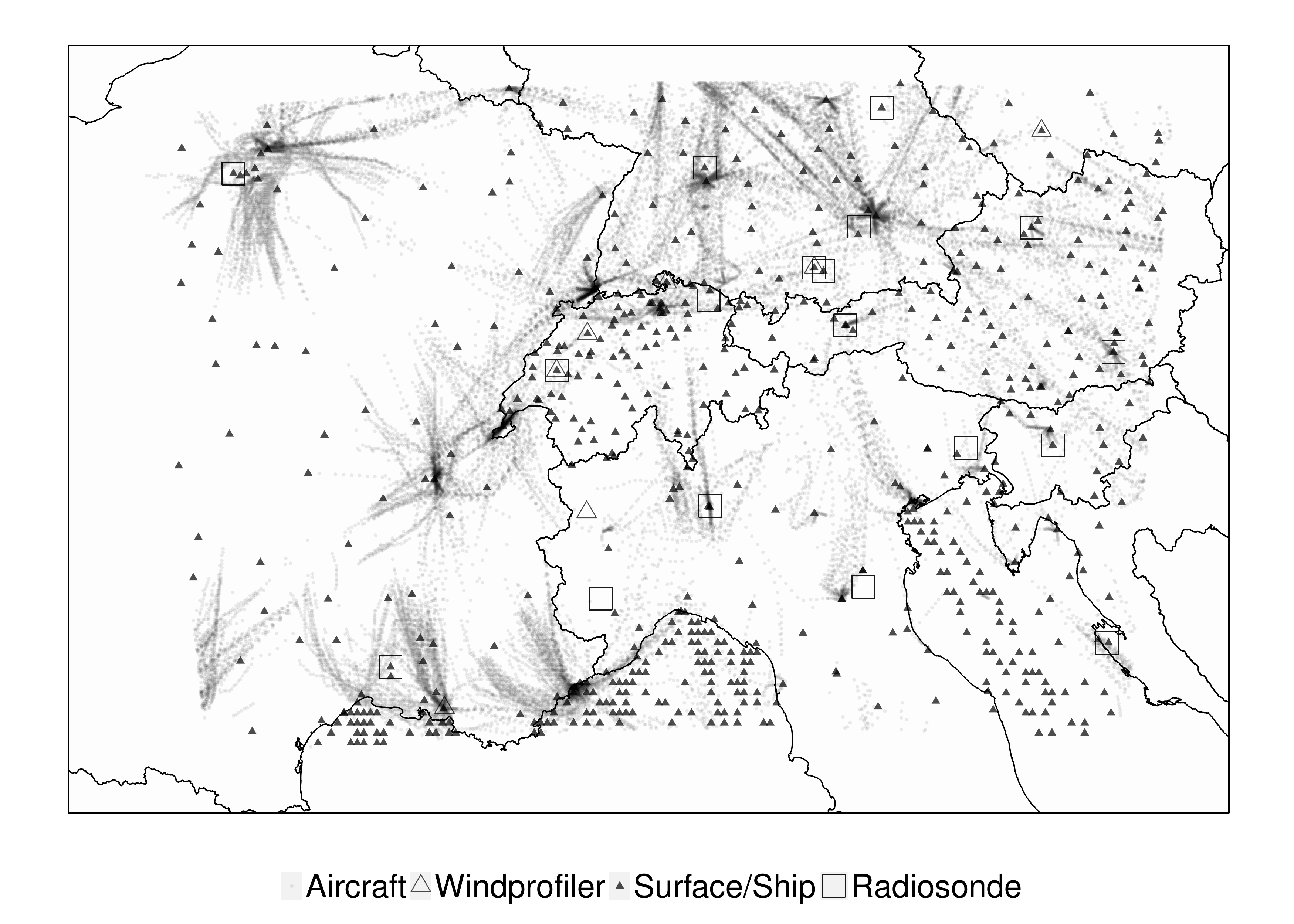}
    \caption{COSMO model domain and geographical distribution of the observations actively assimilated at least once during the 12-day test period.}
    \label{fig:model_domain}
\end{figure}

The MeteoSwiss COSMO implementation covers a geographical domain of central Europe (see \cref{fig:model_domain}) with a horizontal mesh-size of 2.2km and 60 terrain-following vertical levels up to a model top at roughly 22km. 

The reference analysis algorithm is the LETKF based on \citet{hunt_efficient_2007} with a configuration similar to that described in \citet{schraff_2016}. This algorithm is operationally used at MeteoSwiss and serves as a reference for comparisons of the new LETKPF methods.
%\emph{THIS IS NOT WHAT HAPPENS IN REALITY:} For all algorithms, localization is done in observation space using a constant vertical and an adaptive horizontal localization radius in order to ensure a homogeneous effective number of observations throughout the local analyses. 
%The target number of observations in a local batch is chosen to be 100 for an ensemble size of 40. 
%\emph{REPLACE WITH:}
For all algorithms, localization is done in observation space using a constant vertical and horizontal localization radius resulting in a varying effective number of observations being assimilated throughout the analyses. 
A multiplicative, adaptive covariance inflation scheme is used to account for unrepresented model error. 
In the operational MeteoSwiss implementation, additional additive covariance inflation in form of the relaxation to prior perturbation (RTPP) \citep{zhang_2004} method is applied. As RTPP cannot be transferred immediately to the LETKPF we did not use it in this study for comparison reasons.

The KENDA system produces hourly ensemble analyses with 40 members. 
Lateral boundary conditions are taken from the first 40 global ECMWF EPS forecast members interpolated to the COSMO model grid. Ensemble perturbations are then calculated by subtracting the ensemble mean from each member. These perturbations are then added to the latest interpolated ECMWF HRES forecast valid at the same time to build a new ensemble. In order to get a reasonable spread-error relationship at the lateral boundaries, members from an older global ensemble forecast with lead times from +30h to +42h and thus a larger spread are used. The initial ensemble at the start of the test period are obtained from the pre-operational MeteoSwiss KENDA cycle.

The observations used for the experiments are similar to that used operationally at MeteoSwiss: radiosonde (TEMP) temperature, wind and humidity data, wind profiler wind data, surface (SYNOP) and ship surface pressure data and aircraft temperature and wind data. The geographical locations of all observations that were actively assimilated at least once during the 12-day test period are shown in \cref{fig:model_domain}.

The observation error covariance $R$ is assumed to be diagonal with values estimated from innovation statistics following \citet{desroziers_2005} and \citet{li_simultaneous_2009} and are listed in \cref{table:obs_error}. 

% ========================================
% Table with observation errors
% ========================================
\begin{table*}%[ht]
    \centering
    \begin{tabular}{cccc}
        \hline
        Level [hPa]       & Wind [m/s]              & Temperature [K]       & Rel. Humidity [\%] \\
        \hline
        300               & 2.1 / 1.9 / 1.6         & 0.6 / 0.6             & 13.8    \\
        400               & 1.8 / 1.6 / 1.4         & 0.5 / 0.5             & 13.1    \\
        500               & 1.6 / 1.4 / 1.2         & 0.6 / 0.6             & 12.9    \\
        700               & 1.6 / 1.4 / 1.2         & 0.7 / 0.7             & 12.2    \\
        850               & 1.7 / 1.5 / 1.3         & 1.0 / 0.8             & 12.8    \\
        1000              & 1.7 / 1.5 / ~~-~~       & 1.1 / 1.1             & ~~9.3   \\
        \hline
    \end{tabular}
    \caption{Observation errors $\sqrt{\sigma_0^2}$ for wind, temperature and relative humidity at different heights in the atmosphere. The first value is for radiosonde, the second value for aircraft and the third value for wind profiler observations.} % title of Table
    \label{table:obs_error}
\end{table*}

% ========================================
% Figure model domain and obs distribution
% ========================================

\subsubsection{Test period}
The 12-day test period for the experiments from 4 to 16 June 2015 was chosen to include both convective and stratiform precipitation events over the domain of interest. From 4 to 9 June the weather in central Europe was dominated by high pressure systems over northern Europe leading to high surface temperatures and a diurnal cycle of convection over the Alpine Ridge. From 9 to 16 June, a cut-off low west of France and its associated fronts caused several bands of both stratiform and embedded convective precipitation sweeping over the Alps. 

\subsubsection{Assimilation methods}
In all our experiments we compare four assimilation algorithms.
The LETKF is close to the operational setup and serves as a reference. 
%In order to compare the results of the LETKPF with the LETKF, three experiments have been conducted: one-step analysis case study, hourly cycled assimilation, and 24-hour forecasts. 
%Our reference algorithm is the LETKF used operationally.
%Four different algorithms are compared: the LETKF is close to the operational setup and serves as a reference. 
For the LETKPF we test two variants of the algorithm with the  different adaptive $\gamma$ schemes described in \cref{sec:es_algorithm}, which we refer to as LETKPF-ess50 for the scheme targeting a ESS of 50\%, and as LETKPF-minMSE for the scheme minimizing the MSE of the analysis mean. 
The fourth algorithm is the local PF (LPF), defined as our LETKPF with $\gamma$ set to zero.
%As a last method we use the LETKPF with $\gamma$ set to zero everywhere, which is nothing else than a LPF.

%%%%%%%%%%%%%%%%%%%%%%%%%%%%%%%%%%%%%%%%%%%%%%%%%%%%%%%%%%%%%%%%%%%%%
\subsection{Results}
\label{sec:results}
First we show how the LETKPF works and how it differs from the LETKF in a particular one-step analysis case study. Then we present results on the verification of radiosonde observations during the cycling assimilation phase. Finally we look at the 24-hour forecasts and contrast the performance of the different algorithms.

\subsubsection{One-step analysis}

\begin{figure}
    \centering
    \includegraphics[width=\figfactor \textwidth]{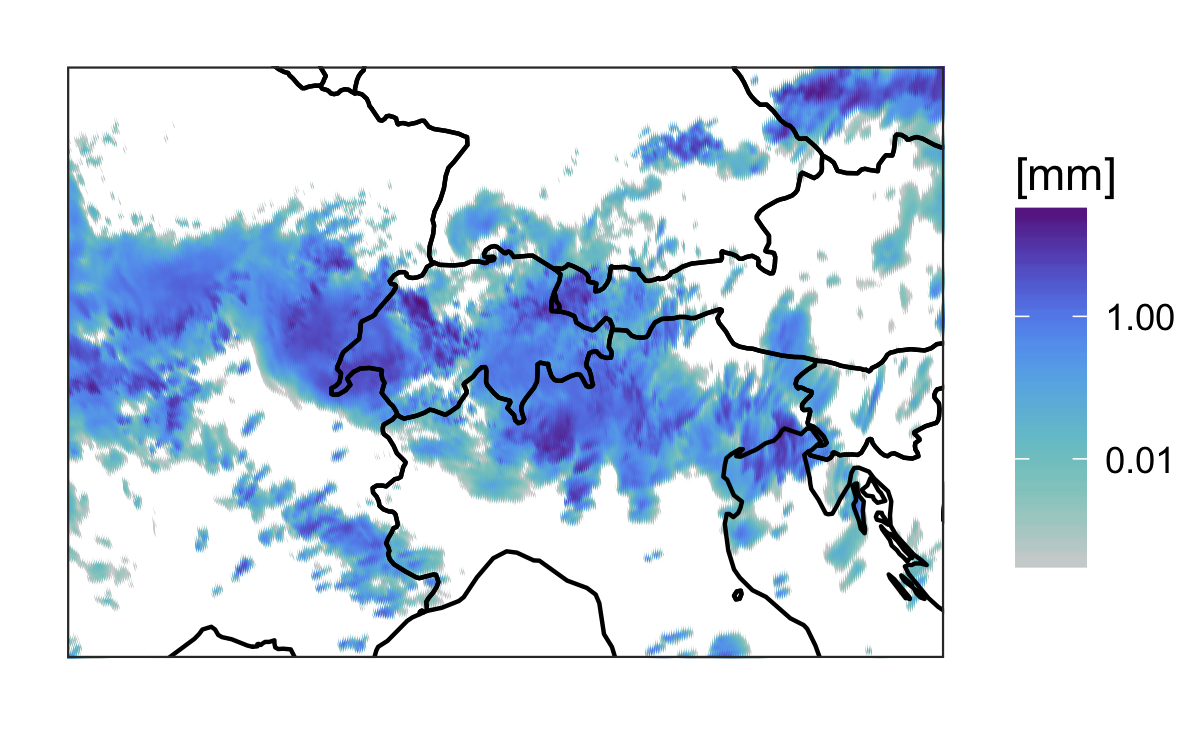}
    \caption{Maps of $\bar x^b$ for total precipitation in [mm] during the last hour before the analysis in the case study.}
    \label{fig:fields}
\end{figure}

We now look in more detail at a one-step analysis on the 14 June at 1700 UTC.
The meteorological situation at analysis time is summarized  in \cref{fig:fields} with the total precipitation of the background mean $\bar x^b$ in [\si{mm}]. A large storm is going through the domain with strong convection happening in many different areas.

\begin{figure}
    \centering
    \includegraphics[width=\figfactor \textwidth]{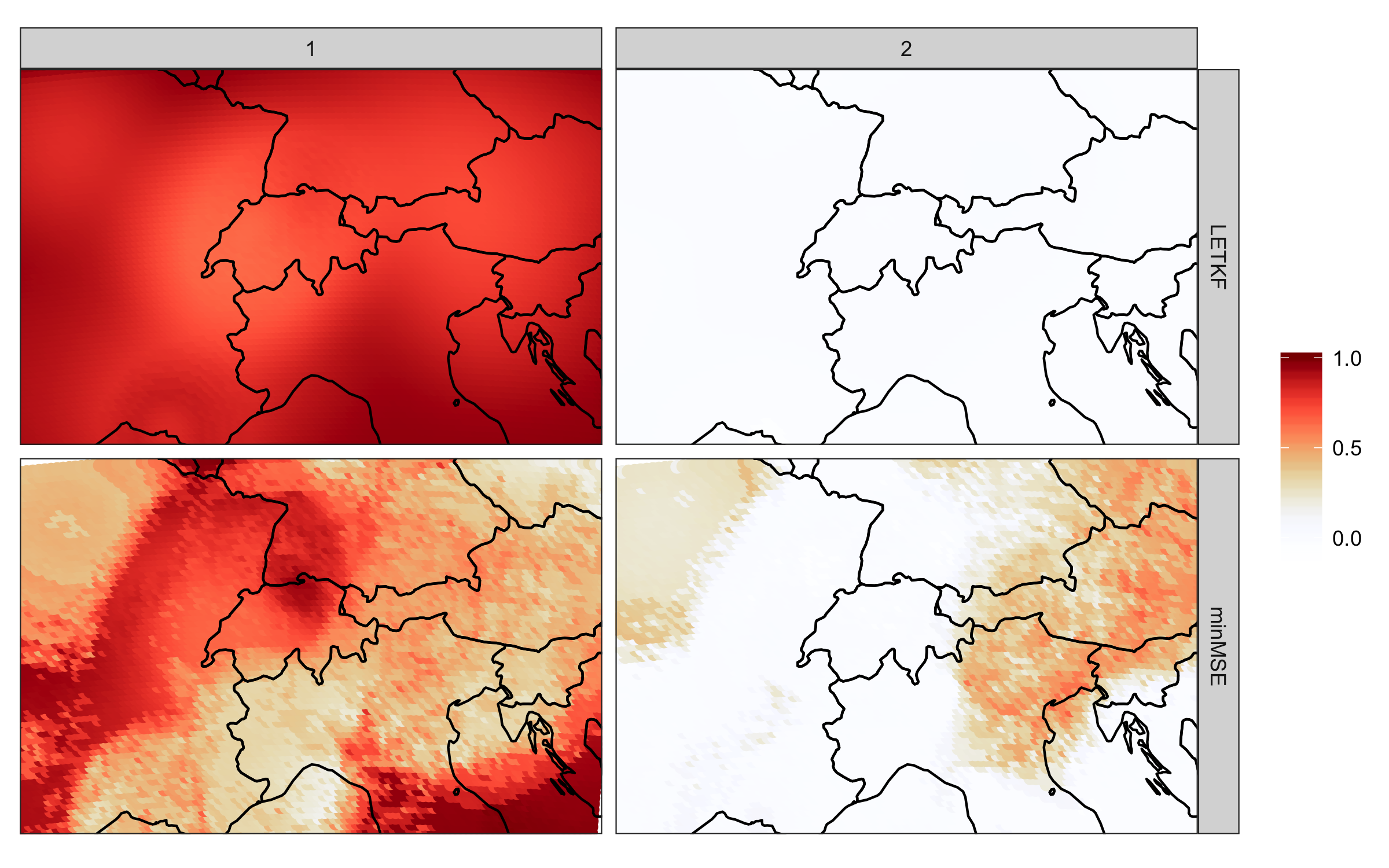}
    \caption{Maps of $W_{i1}$, the contribution of particles $i=1,2$ to the analysis particle $1$ in the lower atmosphere during the case study, when using LETKF in the first row and LETKPF-minMSE in the second. }% The points indicate cities with more than 200,000 inhabitants.}
    \label{fig:wi1}
\end{figure}

To illustrate how the LETKPF differs from the LETKF we look at maps of the analysis weight matrix $W$ 
(to be precise, we look at the values of $W  = \tilde W^{\mu} W^{\alpha} + W^{\epsilon}$, where $\tilde W^{\mu}$ is the left side of \cref{eq:wmueff} only, to remove the effect on the mean and focus on the particle deviations).
%(to be precise we look at the values of $W  = W^{\mu} W^{\alpha} + W^{\epsilon} - 1/k \cdot W^{\mu} W^{\alpha} \bm{1}\bm{1}'$ to remove the effect of the mean and focus on the particle deviations). 
For simplicity, we choose to focus on the contribution of the first two particles to form the analysis particle $x^{a,1}$. These contributions are summarized in the first two elements of the first column of the $W$ matrix, $W_{11}$ and $W_{21}$. Because the analysis is done locally, these values change at every grid point. 
Averaging over the lower atmosphere (pressure larger than \SI{700}{hPa}) we can show the results for different algorithms as maps in \cref{fig:wi1}. 
The particle $x^{a,1}$ is mainly composed of itself -- $x^{b,1}$ -- when the value mapped is close to 1, while it is recomposed from other particles when it is close to 0. When this is the case, other particles are resampled instead and glued together to form the analysis.

In the first row of \cref{fig:wi1} we can see what happens in the case of the LETKF: $x^{a,1}$ is mainly composed of $x^{b,1}$, with the other particles only marginally influencing the analysis through their covariance with $x^{b,1}$. 
In the second row, however, the same maps for the LETKPF-minMSE shows a more interesting behavior: particle $x^{a,1}$ is composed of itself in some areas, for example in North-East France and Switzerland, but in some places $x^{b,1}$ is composed in a large part of $x^{b,2}$  as in Austria and the North-East of Italy, or of other particles not shown here as in the North-West of Italy.
These maps illustrate well how the LETKPF produces an analysis by combining different particles locally, resampling particles where they fit the data well and discarding them where other candidates fit better.

\begin{figure}
    \centering
    \includegraphics[width=\figfactor \textwidth]{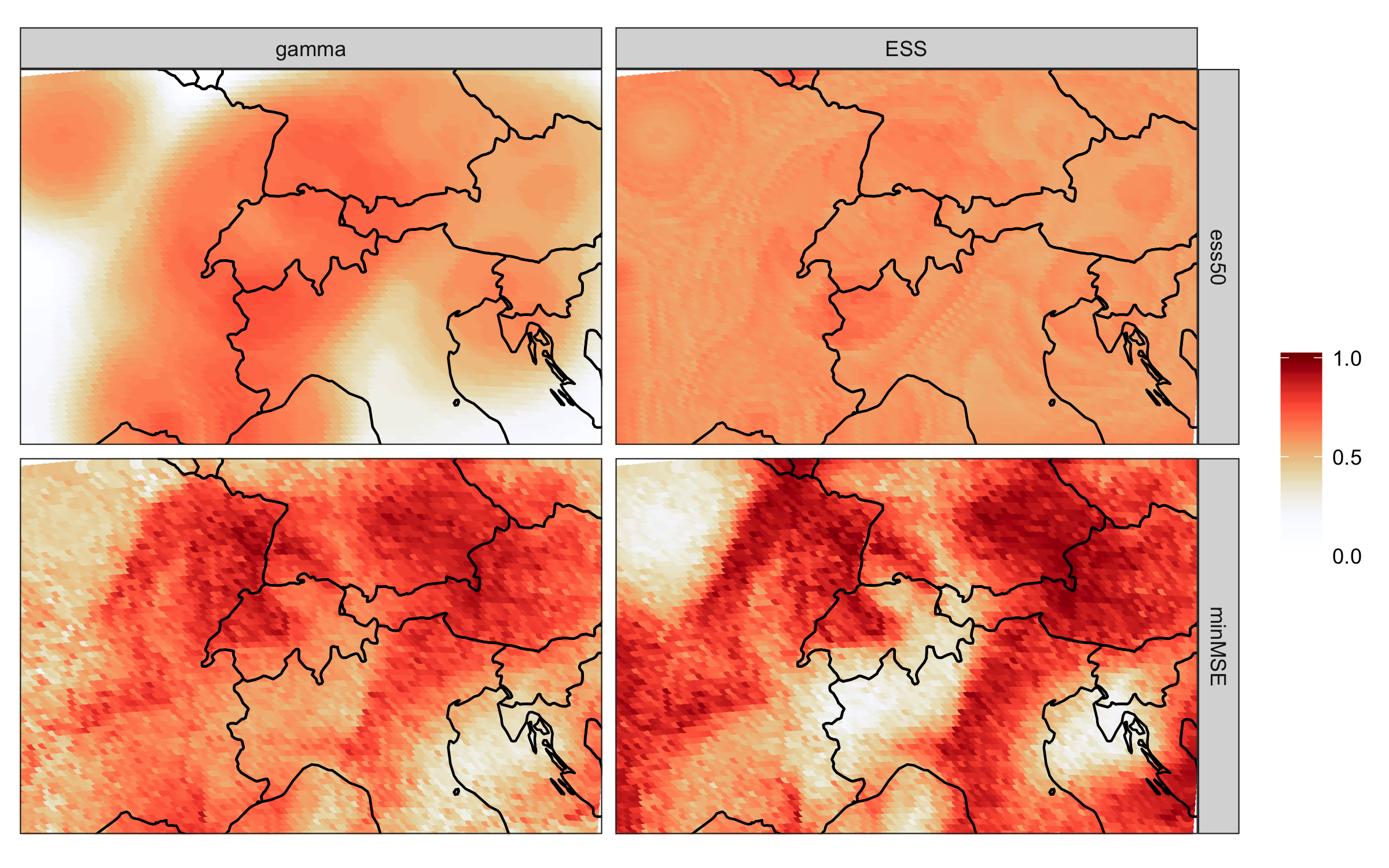}
    \caption{Adaptive choice of $\gamma$ (left panel) and corresponding ESS (right panel) in the lower atmosphere during the case study, with LETKPF-ess50 in the first row and LETKPF-minMSE in the second. }% The points indicate cities with more than 200,000 inhabitants.}
    \label{fig:gam}
\end{figure}

Not only the weights $W$, but also the value of $\gamma$ vary locally. In \cref{fig:gam}, the $\gamma$ chosen in the lower atmosphere with different adaptive criteria is displayed together with the ESS. 
The value of $\gamma$ shows where the algorithm prefers to stay closer to the LETKF (where $\gamma$ is large) and where it chooses an update much closer to the PF (where $\gamma$ is small).
The functional relationship between $\gamma$ and ESS is non-linear and depends locally on the background ensemble distribution and the observations. 
If the ESS is close to 1, little resampling occurs and most particles are reused, while if it is close to 0, a few particles are resampled many times.

The maps in \cref{fig:gam} are quite different for the two algorithms:
the $\gamma$ chosen by the ESS criterion varies less in space, while the $\gamma$ chosen by LETKPF-minMSE has a rougher pattern. 
Both methods agree in some regions of the domain, but in others they make opposite choices, as for example in the region around Paris.  
Unfortunately, there is no ground truth to compare the chosen $\gamma$ with, and one has to rely on the overall performance of a particular algorithm to see if it fared well.
We attempted to find correlations between the choice of $\gamma$ and the meteorological situation, for example by looking at measures of non-Gaussianity, but arrived at no clear result. Furthermore, with the current operational setup the number of observations varies quite a lot in the domain (from 0 to 100), which seems to have a strong influence on the choice of $\gamma$ (see for example in the the region of high-density observations around Paris). 
Further research will be necessary to understand the interplay between the different parameters and the optimal choice of $\gamma$.

\subsubsection{Cycled experiment}

% cycled ASSIMILATION RESULTS 

%Goal: show that new method is more or less equivalent to LETKF with some gains for some aspects and some losses for others.

In order to assess the quality of the analysis during the assimilation, we verify the one-hour-ahead forecast produced by different algorithms against all radiosonde observations.
As error metrics we use the bias of the forecast mean and the CRPS, a strictly proper scoring rule which takes into account both the sharpness and the calibration of the ensemble \citep{gneiting_strictly_2007}. More scores will be considered for the forecast experiment described below, but for the analysis they are sufficient to evaluate the overall performances of the algorithms. 

\begin{figure*}
    \centering
    \includegraphics[width= \figbig \textwidth]{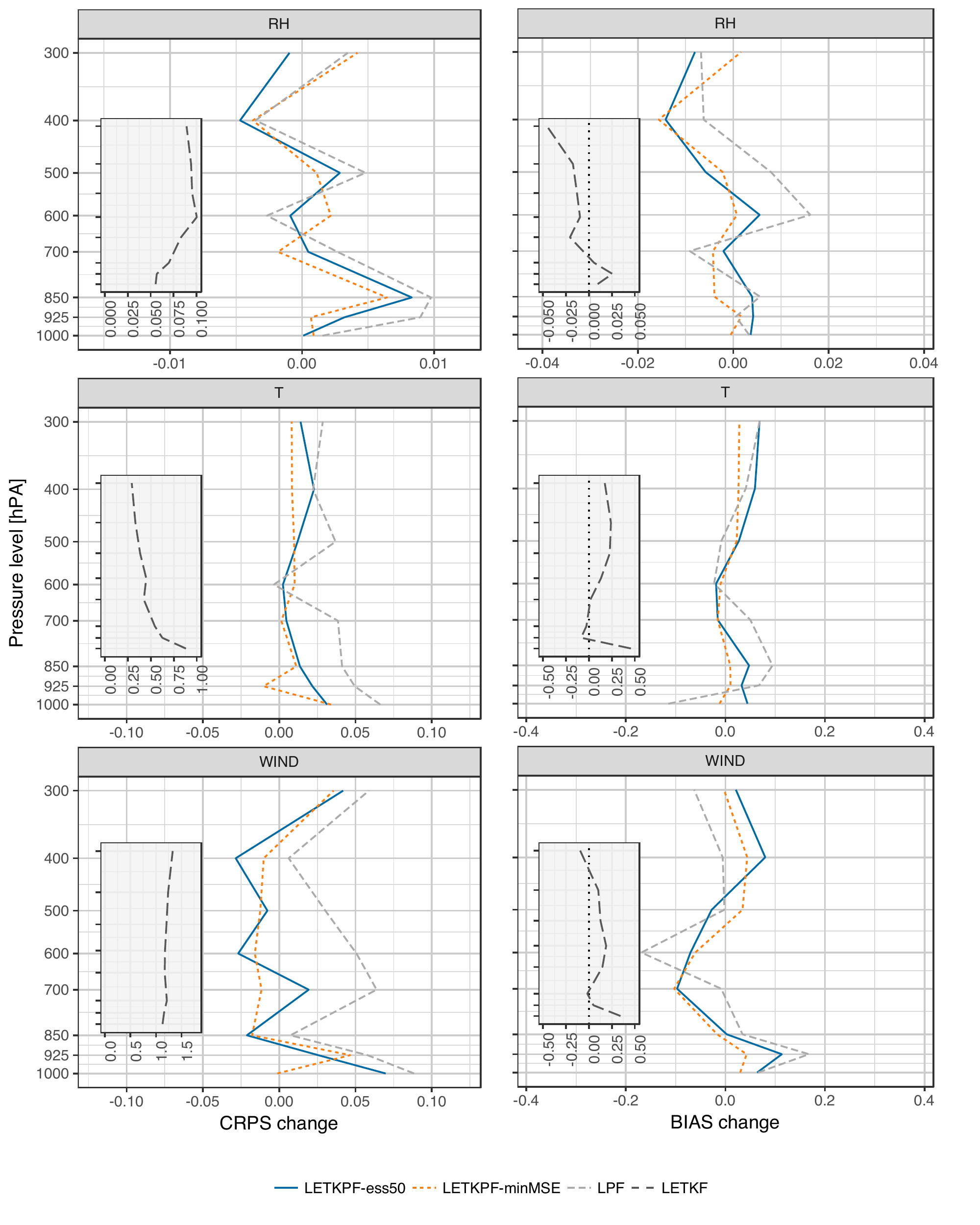}
    \caption{Change in CRPS and bias relative to LETKF analyses during the cycled experiment. More precisely CRPS(LETKPF) - CRPS(LETKF) and $|$bias(LETKPF)$|$ - $|$bias(LETKF)$|$. A negative change indicates a reduction of CRPS or of bias, respectively. Note the different scales on the x-axis. In the small plot is the CRPS and bias of the reference LETKF.}
    \label{fig:profile_changes_cycles}
\end{figure*}

The error metrics are aggregated over the whole period and over different pressure levels. In \cref{fig:profile_changes_cycles}, the difference of the bias and CRPS of the new algorithms with the bias and CRPS of the LETKF are displayed as vertical profiles. For the bias, the difference of the absolute value is displayed, such that for both the bias and the CRPS a negative value indicates an improvement over the LETKF.
In each panel there is a smaller plot included to show the profile of the  LETKF error.

For the relative humidity (RH), the pressure level and the type of method have a strong influence on the CRPS.
It seems that the LETKPFs are worse than the LETKF for the lower atmosphere, but they are sometimes better for the middle and upper atmospheres. There is no clear ranking between the variants of LETKPFs, with LETKPF-minMSE performing best around 700 [hPa] while  LETKPF-ess50 seems better around 400 [hPa]. In terms of bias, we can also see some large gains for the LETKPFs in the middle and upper atmospheres. 

The LETKF predicts temperature (T) better for almost all pressure levels both in terms of CRPS and bias. This comes as no surprise as temperature is the most Gaussian of all the variables. LPF is clearly worse than the other algorithms in terms of CRPS, while it is fares relatively well in terms of bias, particularly at 1000 [hPa]. 

The LETKPFs improve over the LETKF for predicting the wind speed (WIND) at middle to lower atmosphere, as can be seen from the CRPS and bias profiles. The LETKPF-minMSE seems to have the most consistent advantage, if not always the largest. The LPF, on the other hand, has trouble with WIND observations and is the worst method in terms of CRPS while its performance in terms of bias is erratic.

\begin{figure}
    \centering
    \includegraphics[width= \figfactor \textwidth]{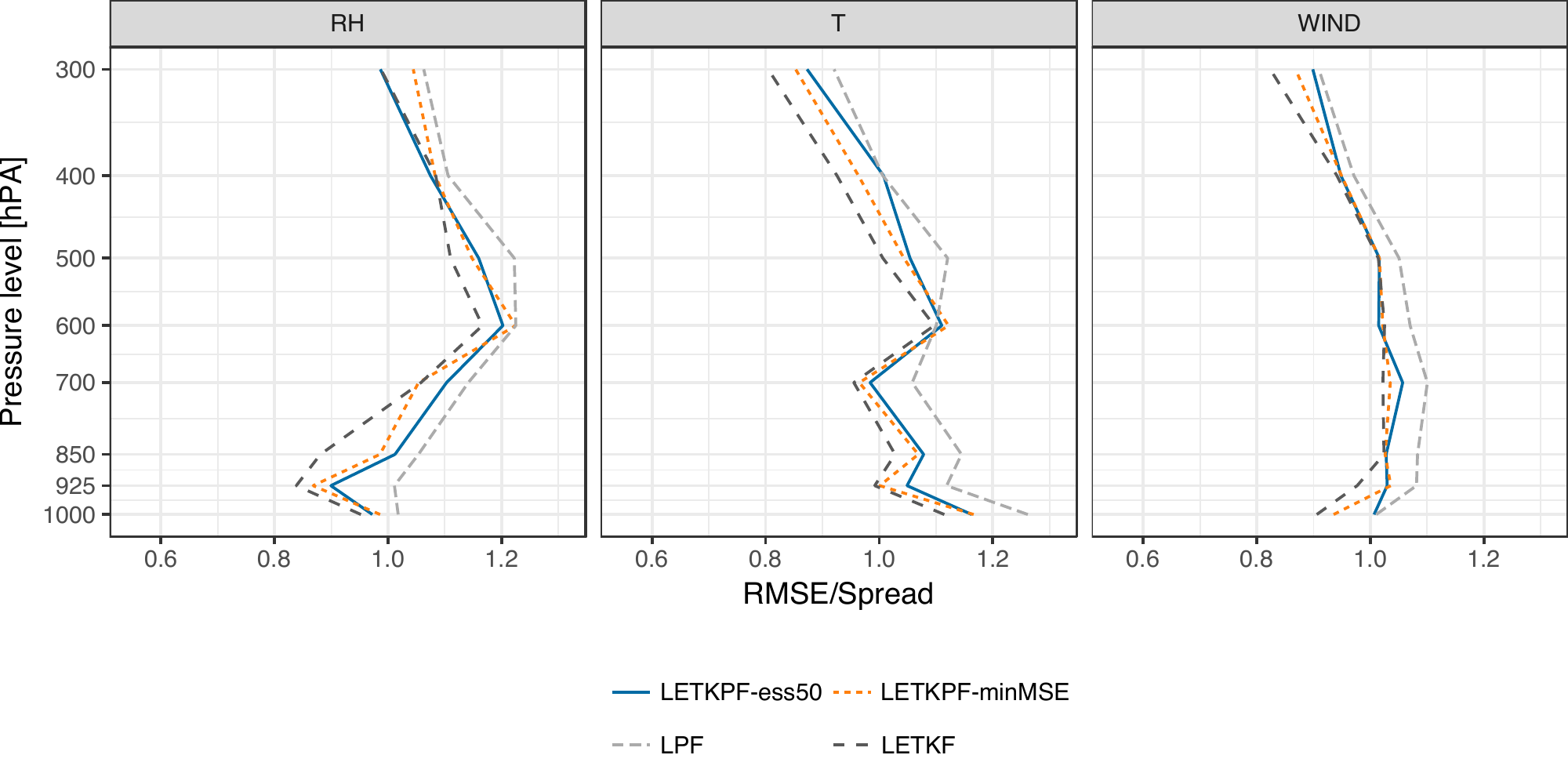}
    \caption{Ratio of RMSE over the spread of the background predictive distribution. A ratio larger than one indicates that the ensemble is too concentrated.}
    \label{fig:rmse-spread}
\end{figure}

In \cref{fig:rmse-spread} we compare the root mean squared error (RMSE) to the spread of the background ensemble, which should be equal if the ensemble is well calibrated (see for example \citet{fortin_why_2014}). 
To take into account the observation error, we actually compare the observed RMSE to the spread of the predictive distribution $\pi^b(y)$. In the case of a diagonal $R$, we can compute this spread squared separately for each observation by adding the variance of the forecast ensemble to the corresponding diagonal element of $R$. We then aggregate by averaging over all observations, and take the square root before comparing to the RMSE.
The profiles in \cref{fig:rmse-spread} show that overall the ensembles are  well calibrated. In terms of relative humidity it seems that the ensembles lack spread in the middle atmosphere, while they are too dispersed in the upper atmosphere in terms of temperature and wind. In general the LETKPFs have a larger ratio than the LETKF, due mainly to a reduction in spread because of resampling.
Better calibration could be achieved in the future by fine tuning of the $R$ matrix and by using refined covariance inflation schemes.

\subsubsection{Forecast experiment}

\begin{figure*}
    \centering
    \includegraphics[width= \figbig \textwidth]{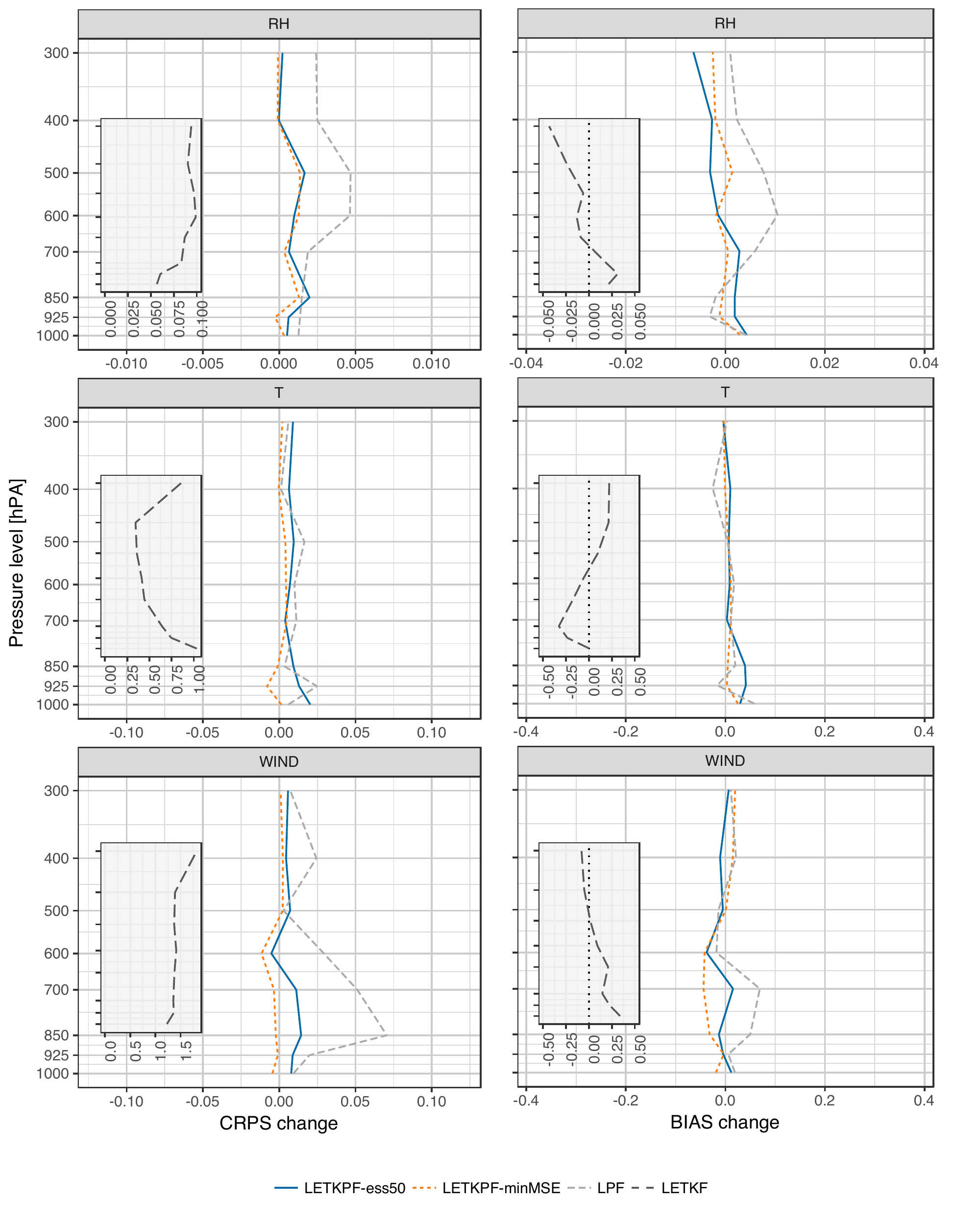}
    \caption{Change in forecast CRPS and bias relative to LETKF-driven forecasts, similar to \cref{fig:profile_changes_cycles}. The scores of all forecast lead times were aggregated. A negative change indicates a reduction of CRPS or of bias, respectively. Note the different scales on the x-axis. In the small plot is the CRPS and bias of the reference LETKF.}
    \label{fig:profile_changes_fcst}
\end{figure*}

Twice a day, at 0000 and 1200 UTC, a 24-hour forecast was launched from the current analysis ensemble. 
In \cref{fig:profile_changes_fcst} we look at the CRPS and bias of predicting radiosonde observations averaged over the whole domain and the whole forecast horizon (i.e. the scores of all forecast lead times were aggregated to one single score),
similar to the cycled experiments in \cref{fig:profile_changes_cycles}. 
The absolute CRPS is usually larger in the forecast than in the cycled experiment, with the strongest growth in the upper atmosphere for the temperature and wind variables. 
However, the differences between the methods are much less pronounced than during the analysis and disappear almost completely at the end of the 24-hour forecast. 
The LPF is clearly worse than the other algorithms, particularly in terms of relative humidity and wind. For the relative humidity and temperature the LETKF is generally slightly better, while among the LETKPFs the LETKPF-minMSE is the best performer and even beats the LETKF for the wind variables at most levels.

\begin{figure*}
    \centering
    \includegraphics[width= \figbig \textwidth]{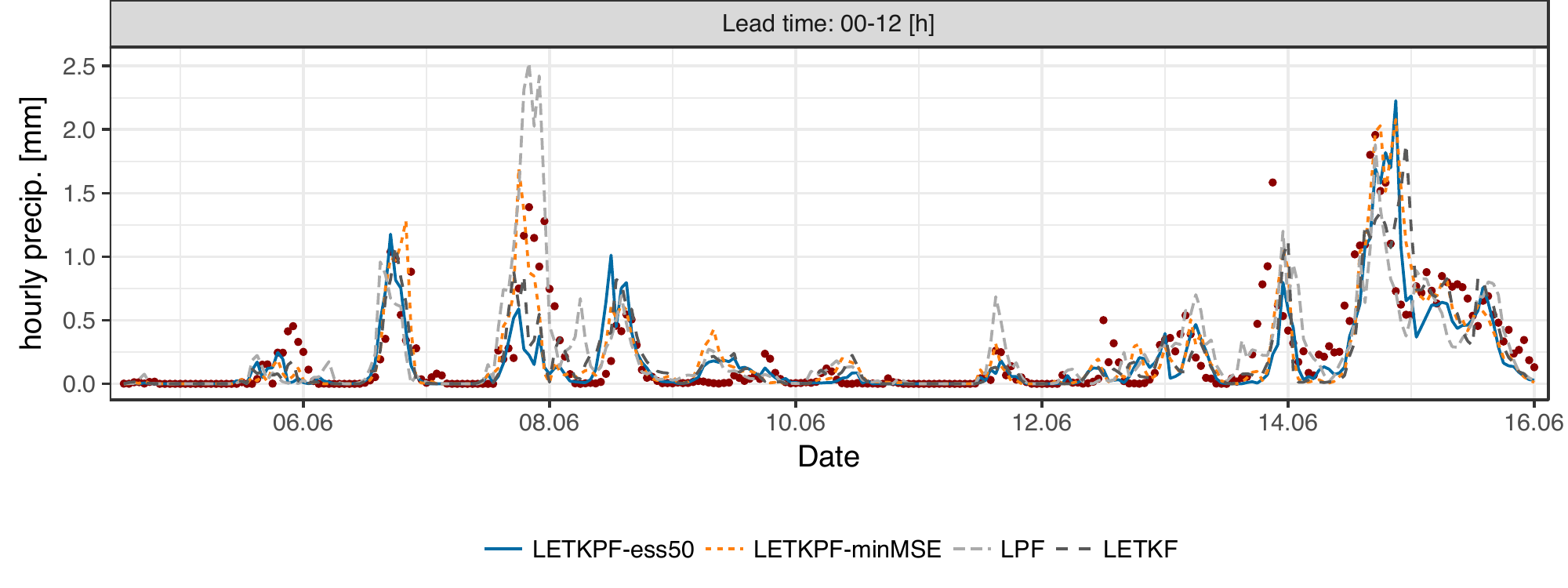}
    \caption{Time evolution of the ensemble mean (various lines) compared to the observations (dots) of hourly precipitations from 121 Swiss SYNOP stations.}
    \label{fig:fcst_obsts}
\end{figure*}

More relevant for the forecast users, we now look at the hourly precipitation recorded at 121 stations over the Swiss domain (SYNOP data).  
In \cref{fig:fcst_obsts} we can see the evolution of the ensemble forecast means (the first 12 lead time hours of all forecasts are chunked together to build a continuous time series) over the whole period as compared to the actual observations (dots). It is interesting to notice how the different algorithms coincide most of the time but differ substantially for some events. For example, around the 8 June a large precipitation event is best predicted by LETKPF-minMSE forecasts, while the LPF forecasts overestimate it, and the other method underestimate it. At other times, all methods seem to miss or produce spurious events.

\begin{figure}
    \centering
    \includegraphics[width= \figfactor \textwidth]{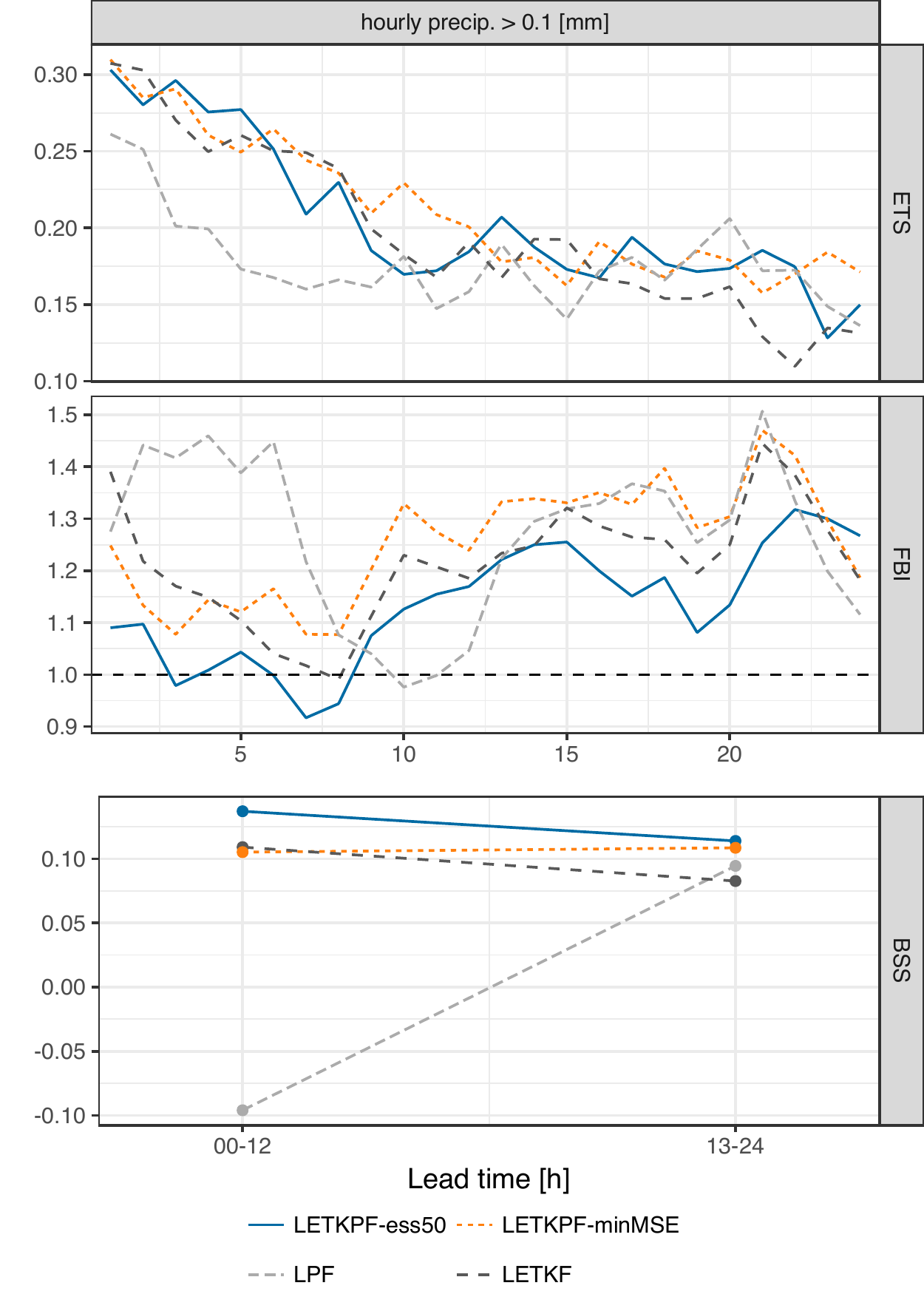}
    \caption{Evolution of various scores for predicting hourly precipitation larger than 0.1 [mm] as a function of lead time during forecast, and aggregated over all forecasts of the period under consideration. Reference observations are accumulated hourly precipitation from 121 Swiss SYNOP stations. For ETS and BSS the higher the better (maximum 1), and for FBI the closer to 1 the better. Because of system constraints, the BSS is aggregated every 12 hours, whereas the other scores are aggregated hourly.}
    \label{fig:fcst_scores}
\end{figure}

The evolution of the skills of the ensemble to predict hourly precipitation larger than 0.1 [mm] as a function of lead time is illustrated in \cref{fig:fcst_scores}, where we see the equitable threat score (ETS), the frequency bias index (FBI) and the Brier skill score (BSS) of the forecast ensembles. In terms of ETS, the LETKPFs and the LETKF are more or less equivalent, with some lead time where one or another is better. The LPF on the other hand is clearly worse during the first 12 hours of forecast but then stabilizes. The FBI plot shows that all methods tend to overforecast the event, while the LETKPF-ess50 has the best overall performance. For the BSS, the Brier score normalized by the climatology forecast score (as computed from the test period), the LETKPF-ess50 is again the best performer, while the LPF has no skill in the first half of the forecast but reaches similar level to the others in the second half.

\begin{figure*}
    \centering
    \includegraphics[width= \figbig \textwidth]{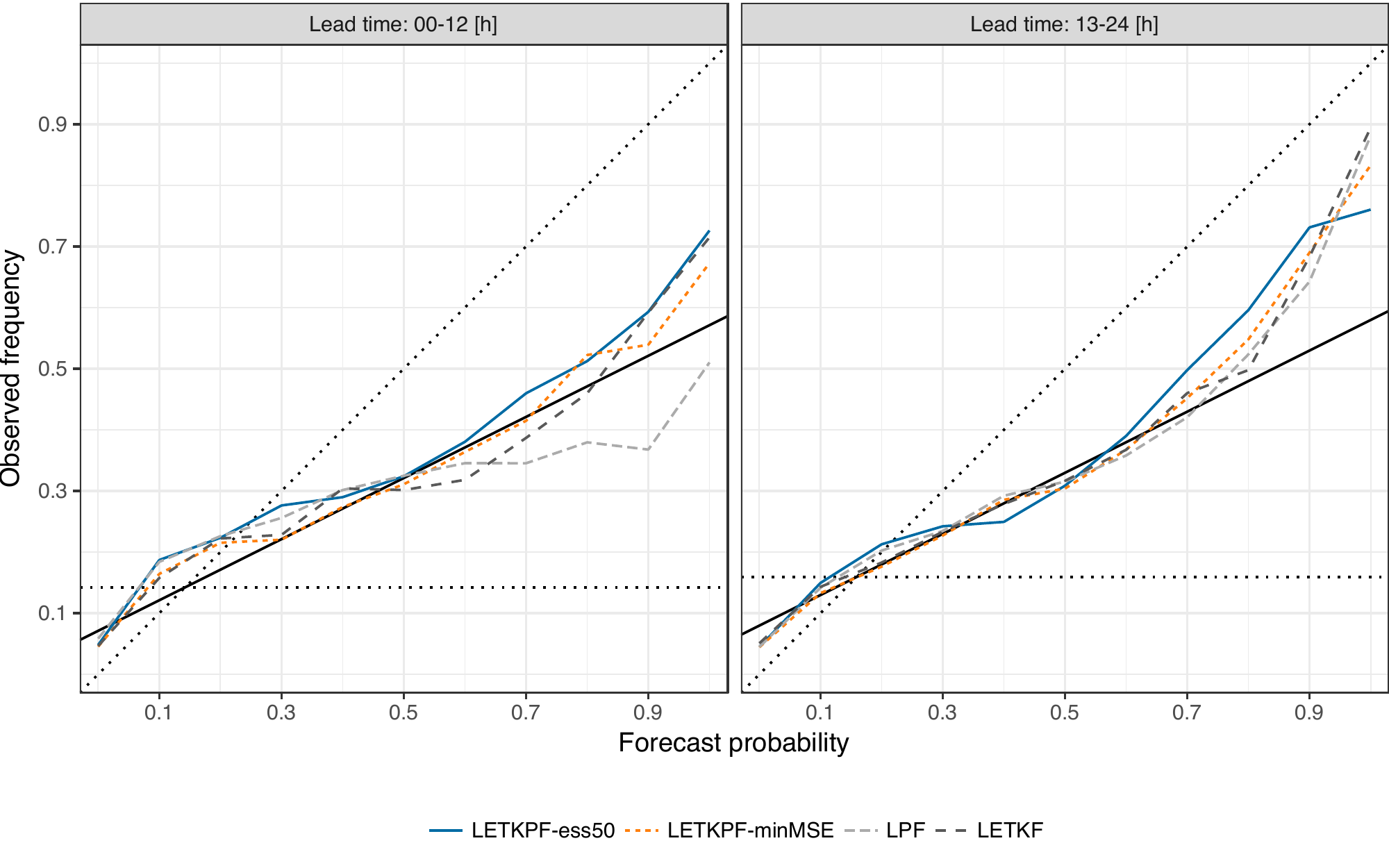}
    \caption{Reliability diagram for predicting more than 0.1 [mm] of hourly precipitation in the first 12 hours (left panel) and the second 12 hours (right panel). The solid black line indicates no skill while the diagonal is for perfect reliability. }
    \label{fig:fcst_reldiag}
\end{figure*}

The calibration of the methods is shown in the reliability diagrams of \cref{fig:fcst_reldiag}. One can see that all algorithms have some skill except maybe the LPF during the first 12 hours of forecast. The LETKPF-ess50 is once again the best performer and the LETKF is generally less well calibrated than the LETKPFs, but the differences are small and depend on the forecast probabilities. The rank histograms indicate an overall positive bias, but no particular differences between the methods (not shown).

\subsubsection{Discussion}

The results of the cycled and forecast experiments show that the  LETKPFs perform similarly to the LETKF. 
The new algorithms bring some improvements for some variables at some pressure levels -- for example for wind in the middle and upper atmosphere -- but they also perform worse in other cases.
As expected, these improvements over the LETKF occur for the most non-Gaussian variables, while for Gaussian variables like temperature the LETKF is usually better. 
During the forecast in particular, the LETKPFs show some benefit in predicting hourly precipitation, which is a highly non-Gaussian variable. The better ability of the EnKPF to deal with rain fields confirm previous results with a toy model of cumulus convection \citep{robert_localizing_2017}.

The LPF is surprisingly not as bad as one could expect given its simplicity, which shows that localizing the PF is a viable strategy, but the ability to combine it with the LETKF seems to bring clear advantages. 
However, the question of which criterion to use for choosing the proportion of PF and  of LETKF in the analysis is still not clear from the empirical results. The LETKPF-minMSE seems to be slightly better for the model variables (temperature, relative humidity and wind), but the LETKPF-ess50 typically performs better for forecasting hourly precipitation. 

These results are promising and indicate that the LETKPF can be used in practice. However, further experiments should be conducted with longer periods and during different meteorological situations. 
%It will also be important to quantify the uncertainty in the error estimates, which we did not do here because of system constraints. 

%%%%%%%%%%%%%%%%%%%%%%%%%%%%%%%%%%%%%%%%%%%%%%%%%%%%%%%%%%%%%%%%%%%%%
\section{Summary and conclusions}
\label{sec:conclusion}

High-dimensional non-Gaussian filtering problems, such as encountered in convective scale data assimilation, call for the development of new algorithms. In the present paper we proposed the LETKPF, which builds on the EnKPF to make it more efficient and applicable in practice. In particular, we reformulated the whole algorithm in ensemble space and derived a deterministic scheme such that it now has the ETKF instead of the stochastic EnKF as a limiting case.
The same approach as that of the LETKF was taken for localizing the algorithm, with a few additional steps to deal with the PF nature of the analysis. While this may not be the optimal localization strategy, it is widely used in practice and made the implementation in the existing framework feasible.
Furthermore, a new criterion for choosing the proportion of analysis to be done with the PF and the ETKF was proposed based on the idea of minimizing the predictive MSE.

The new algorithm was implemented in the COSMO data assimilation framework and tested on a 12-day period of hourly assimilation in a region surrounding Switzerland. These experiments showed that the newly proposed algorithm is applicable in practice and can perform similarly to the LETKF, which is the algorithm used operationally at MeteoSwiss.
In particular, the LETKPF brings some remarkable improvements for non-Gaussian variables such as wind and hourly precipitation.
These results are promising and we hope that they will stimulate further experiments and research with the LETKPF and other types of localized hybrid algorithms.

In the present study, we have relied on the setup used for the LETKF, but some questions concerning the particularity of the LETKPF -- or more generally any hybrid algorithm -- need to be further investigated. The optimal choice of $\gamma$ is still poorly understood and the experimental results were not conclusive, showing that both proposed methods work better in some situations.
The alternatives discussed in 
\cref{sec:adaptive_gamma} might be promising and could be tested in practice if efficient implementations are found. 
In general, it would be of great interest to better understand the interplay between the optimal choice of $\gamma$ and the non-Gaussianity of the  distribution, the number of observations assimilated, the model error, etc. As we have seen in the experiments, the choice of $\gamma$ should certainly vary for every grid point, as different situations call for different decisions. One could push this idea further and choose a different $\gamma$ for different types of observations or even for different model variables. For example, one could imagine using a $\gamma$ close to 1 for temperature while using a small $\gamma$ for wind or relative humidity. %It is not clear how we could do that but it should be of interest to the practitioner. 

Another aspect that should be explored further is how to control the ensemble spread for the LETKPF. 
%As mentioned in \cref{sec:kenda_system}, RTPP is not directly usable for the LETKPF. 
Among other means, to do so the LETKF relies on covariance inflation (multiplicative and additive) and RTPP. 
However, both of these methods derive their rationale from the idea that the analysis consists in moving a little bit each particle such that the new ensemble has a correct mean and covariance. RTPP controls the loss of spread by recombining each analysis particle with its corresponding background particle, while covariance inflation somehow increases the analysis ensemble covariance. Because the LETKPF analysis consists partly in resampling particles, one cannot just transpose these techniques blindly. One obvious solution to this issue would be to work with the mixture representation of the analysis and control the spread by adding more covariance to the mixture components. Similarly, for RTPP one could use the idea of combining the analysis particle with the background ensemble, but by taking into account the resampling step of the analysis.

\section*{Acknowledgments}
We would like to thank the data assimilation team of the Deutscher Wetter Dienst, in particular Roland Potthast for his helpful comments and guidance, and Andreas Rhodin for his support with the implementation of the code into the COSMO assimilation framework. 
All simulations in this study have been conducted at the Swiss National Supercomputing Centre.

\bibliographystyle{wileyqj}
%\bibliography{references}
\bibliography{alref}

\begin{thebibliography}{35}
\providecommand{\natexlab}[1]{#1}
\providecommand{\url}[1]{\texttt{#1}}
\providecommand{\urlprefix}{URL }
\expandafter\ifx\csname urlstyle\endcsname\relax
  \providecommand{\doi}[1]{doi:\discretionary{}{}{}#1}\else
  \providecommand{\doi}{doi:\discretionary{}{}{}\begingroup
  \urlstyle{rm}\Url}\fi

\bibitem[{Ades and van Leeuwen(2013)}]{ades_exploration_2013}
Ades M, van Leeuwen PJ. 2013. An exploration of the equivalent weights particle
  filter. \emph{Quarterly Journal of the Royal Meteorological Society}
  \textbf{139}(672): 820--840, \doi{10.1002/qj.1995}.

\bibitem[{Baldauf \emph{et~al.}(2011)Baldauf, Seifert, F{\"{o}}rstner,
  Majewski, Raschendorfer and Reinhardt}]{baldauf_2011}
Baldauf M, Seifert A, F{\"{o}}rstner J, Majewski D, Raschendorfer M, Reinhardt
  T. 2011. Operational convective-scale numerical weather prediction with the
  {COSMO} model: Description and sensitivities. \emph{Monthly Weather Review}
  \textbf{139}(12): 3887--3905, \doi{10.1175/MWR-D-10-05013.1}.

\bibitem[{Bartels and Stewart(1972)}]{bartels_solution_1972}
Bartels RH, Stewart GW. 1972. Solution of the matrix equation {AX}+ {XB}= {C}
  [{F}4]. \emph{Communications of the ACM} \textbf{15}(9): 820--826.

\bibitem[{Bauer \emph{et~al.}(2015)Bauer, Thorpe and Brunet}]{bauer_quiet_2015}
Bauer P, Thorpe A, Brunet G. 2015. The quiet revolution of numerical weather
  prediction. \emph{Nature} \textbf{525}(7567): 47--55,
  \doi{10.1038/nature14956}.

\bibitem[{Carpenter \emph{et~al.}(1999)Carpenter, Clifford and
  Fearnhead}]{carpenter_improved_1999}
Carpenter J, Clifford P, Fearnhead P. 1999. Improved particle filter for
  nonlinear problems. \emph{IEE Proceedings-Radar, Sonar and Navigation}
  \textbf{146}(1): 2--7, \doi{10.1049/ip-rsn:19990255}.

\bibitem[{Chustagulprom \emph{et~al.}(2016)Chustagulprom, Reich and
  Reinhardt}]{chustagulprom_hybrid_2016}
Chustagulprom N, Reich S, Reinhardt M. 2016. A hybrid ensemble transform
  particle filter for nonlinear and spatially extended dynamical systems.
  \emph{SIAM/ASA Journal on Uncertainty Quantification} :
  592--608\doi{10.1137/15M1040967}.

\bibitem[{Crisan(2001)}]{crisan_particle_2001}
Crisan D. 2001. Particle filters --- {A} theoretical perspective. In:
  \emph{Sequential {Monte} {Carlo} {Methods} in {Practice}}, Doucet A, Freitas
  Nd, Gordon N\ (eds), Statistics for {Engineering} and {Information}
  {Science}, Springer New York, pp. 17--41, \doi{10.1007/978-1-4757-3437-9_2}.

\bibitem[{de~Wiljes \emph{et~al.}(2016)de~Wiljes, Acevedo and
  Reich}]{de_wiljes_second-order_2016}
de~Wiljes J, Acevedo W, Reich S. 2016. Second-order accurate ensemble transform
  particle filters. \emph{arXiv:1608.08179} .

\bibitem[{Desroziers \emph{et~al.}(2005)Desroziers, Berre, Chapnik and
  Poli}]{desroziers_2005}
Desroziers G, Berre L, Chapnik B, Poli P. 2005. Diagnosis of observation,
  background and analysis-error statistics in observation space.
  \emph{Quarterly Journal of the Royal Meteorological Society}
  \textbf{131}(613): 3385--3396, \doi{10.1256/qj.05.108}.

\bibitem[{Doucet \emph{et~al.}(2001)Doucet, Freitas and
  Gordon}]{doucet_smc_2001}
Doucet A, Freitas N, Gordon N (eds). 2001. \emph{Sequential {Monte} {Carlo}
  {Methods} in {Practice}}. Springer New York: New York, NY,
  \doi{10.1007/978-1-4757-3437-9}.

\bibitem[{Evensen(1994)}]{evensen_sequential_1994}
Evensen G. 1994. Sequential data assimilation with a nonlinear
  quasi-geostrophic model using {Monte} {Carlo} methods to forecast error
  statistics. \emph{Journal of Geophysical Research: Oceans} \textbf{99}(C5):
  10\,143--10\,162, \doi{10.1029/94JC00572}.

\bibitem[{Evensen(2009)}]{evensen_data_2009}
Evensen G. 2009. \emph{Data {Assimilation}: {The} {Ensemble} {Kalman}
  {Filter}}. Springer Science \& Business Media,
  \doi{10.1007/s10236-003-0036-9}.

\bibitem[{Fortin \emph{et~al.}(2014)Fortin, Abaza, Anctil and
  Turcotte}]{fortin_why_2014}
Fortin V, Abaza M, Anctil F, Turcotte R. 2014. Why should ensemble spread match
  the rmse of the ensemble mean? \emph{Journal of Hydrometeorology}
  \textbf{15}(4): 1708--1713, \doi{10.1175/JHM-D-14-0008.1}.

\bibitem[{Frei and K{\"{u}}nsch(2013)}]{frei_enkpf_2013}
Frei M, K{\"{u}}nsch HR. 2013. Bridging the ensemble {Kalman} and particle
  filters. \emph{Biometrika} : 781--800\doi{10.1093/biomet/ast020}.

\bibitem[{Gneiting and Raftery(2005)}]{gneiting_weather_2005}
Gneiting T, Raftery AE. 2005. Weather forecasting with ensemble methods.
  \emph{Science} \textbf{310}(5746): 248--249, \doi{10.1126/science.1115255}.

\bibitem[{Gneiting and Raftery(2007)}]{gneiting_strictly_2007}
Gneiting T, Raftery AE. 2007. Strictly proper scoring rules, prediction, and
  estimation. \emph{Journal of the American Statistical Association}
  \textbf{102}(477): 359--378, \doi{10.1198/016214506000001437}.

\bibitem[{Gordon \emph{et~al.}(1993)Gordon, Salmond and
  Smith}]{gordon_novel_1993}
Gordon N, Salmond D, Smith A. 1993. Novel approach to nonlinear/non-{Gaussian}
  {Bayesian} state estimation. \emph{Radar and Signal Processing, IEE
  Proceedings F} \textbf{140}(2): 107--113, \doi{10.1049/ip-f-2.1993.0015}.

\bibitem[{Harnisch and Keil(2015)}]{harnisch_initial_2015}
Harnisch F, Keil C. 2015. Initial conditions for convective-scale ensemble
  forecasting provided by ensemble data assimilation. \emph{Monthly Weather
  Review} \textbf{143}(5): 1583--1600, \doi{10.1175/MWR-D-14-00209.1}.

\bibitem[{Hunt \emph{et~al.}(2007)Hunt, Kostelich and
  Szunyogh}]{hunt_efficient_2007}
Hunt BR, Kostelich EJ, Szunyogh I. 2007. Efficient data assimilation for
  spatiotemporal chaos: {A} local ensemble transform {Kalman} filter.
  \emph{Physica D: Nonlinear Phenomena} \textbf{230}(1-2): 112--126,
  \doi{10.1016/j.physd.2006.11.008}.

\bibitem[{K{\"{u}}nsch(2005)}]{kunsch_recursive_2005}
K{\"{u}}nsch HR. 2005. Recursive {Monte} {Carlo} filters: {Algorithms} and
  theoretical analysis. \emph{The Annals of Statistics} \textbf{33}(5):
  1983--2021, \doi{10.1214/009053605000000426}.

\bibitem[{Lancaster and Rodman(1995)}]{lancaster_algebraic_1995}
Lancaster P, Rodman L. 1995. \emph{Algebraic {Riccati} {Equations}}. Clarendon
  Press.

\bibitem[{Li \emph{et~al.}(2009)Li, Kalnay and Miyoshi}]{li_simultaneous_2009}
Li H, Kalnay E, Miyoshi T. 2009. Simultaneous estimation of covariance
  inflation and observation errors within an ensemble {Kalman} filter.
  \emph{Quarterly Journal of the Royal Meteorological Society}
  \textbf{135}(639): 523--533, \doi{10.1002/qj.371}.

\bibitem[{Liu(1996)}]{liu_metropolized_1996}
Liu JS. 1996. Metropolized independent sampling with comparisons to rejection
  sampling and importance sampling. \emph{Statistics and Computing}
  \textbf{6}(2): 113--119, \doi{10.1007/BF00162521}.

\bibitem[{Pitt and Shephard(1999)}]{pitt_filtering_1999}
Pitt MK, Shephard N. 1999. Filtering via simulation: Auxiliary particle
  filters. \emph{Journal of the American Statistical Association}
  \textbf{94}(446): 590--599, \doi{10.1080/01621459.1999.10474153}.

\bibitem[{Poterjoy(2016)}]{poterjoy_localized_2016}
Poterjoy J. 2016. A localized particle filter for high-dimensional nonlinear
  systems. \emph{Monthly Weather Review} \textbf{144}(1): 59--76,
  \doi{10.1175/MWR-D-15-0163.1}.

\bibitem[{Poterjoy and Anderson(2016)}]{poterjoy_efficient_2016}
Poterjoy J, Anderson JL. 2016. Efficient assimilation of simulated observations
  in a high-dimensional geophysical system using a localized particle filter.
  \emph{Monthly Weather Review} \textbf{144}(5): 2007--2020,
  \doi{10.1175/MWR-D-15-0322.1}.

\bibitem[{Rebeschini and Handel(2015)}]{rebeschini_can_2015}
Rebeschini P, Handel Rv. 2015. Can local particle filters beat the curse of
  dimensionality? \emph{The Annals of Applied Probability} \textbf{25}(5):
  2809--2866, \doi{10.1214/14-AAP1061}.

\bibitem[{Reich(2013)}]{reich_nonparametric_2013}
Reich S. 2013. A nonparametric ensemble transform method for {Bayesian}
  inference. \emph{SIAM Journal on Scientific Computing} \textbf{35}(4):
  A2013--A2024, \doi{10.1137/130907367}.

\bibitem[{{Robert} and {K{\"u}nsch}(2017)}]{robert_localization_2016}
{Robert} S, {K{\"u}nsch} HR. 2017. Localization in high-dimensional {Monte}
  {Carlo} filtering. In: \emph{Bayesian Statistics in Action}, vol. 194,
  Argiento R, Lanzarone E, Villalobos IA, Mattei A\ (eds), ch.~8, Springer
  Proceedings in Mathematics and Statistics, pp. 79--89.

\bibitem[{Robert and K{\"{u}}nsch(2017)}]{robert_localizing_2017}
Robert S, K{\"{u}}nsch HR. 2017. Localizing the ensemble {Kalman} particle
  filter. \emph{Tellus A: Dynamic Meteorology and Oceanography} \textbf{69}(1):
  1--14, \doi{10.1080/16000870.2017.1282016}.

\bibitem[{Schraff \emph{et~al.}(2016)Schraff, Reich, Rhodin, Schomburg,
  Stephan, Peri{\'a}{\~n}ez and Potthast}]{schraff_2016}
Schraff C, Reich H, Rhodin A, Schomburg A, Stephan K, Peri{\'a}{\~n}ez A,
  Potthast R. 2016. Kilometre-scale ensemble data assimilation for the cosmo
  model kenda. \emph{Quarterly Journal of the Royal Meteorological Society}
  \textbf{142}(696): 1453--1472, \doi{10.1002/qj.2748}.

\bibitem[{Snyder \emph{et~al.}(2008)Snyder, Bengtsson, Bickel and
  Anderson}]{snyder_obstacles_2008}
Snyder C, Bengtsson T, Bickel P, Anderson J. 2008. Obstacles to
  {High}-{Dimensional} {Particle} {Filtering}. \emph{Monthly Weather Review}
  \textbf{136}(12): 4629--4640, \doi{10.1175/2008MWR2529.1}.

\bibitem[{Snyder \emph{et~al.}(2015)Snyder, Bengtsson and
  Morzfeld}]{snyder_performance_2015}
Snyder C, Bengtsson T, Morzfeld M. 2015. Performance bounds for particle
  filters using the optimal proposal. \emph{Monthly Weather Review}
  \textbf{143}(11): 4750--4761, \doi{10.1175/MWR-D-15-0144.1}.

\bibitem[{van Leeuwen(2010)}]{van_leeuwen_nonlinear_2010}
van Leeuwen PJ. 2010. Nonlinear data assimilation in geosciences: an extremely
  efficient particle filter. \emph{Quarterly Journal of the Royal
  Meteorological Society} \textbf{136}(653): 1991--1999, \doi{10.1002/qj.699}.

\bibitem[{Zhang \emph{et~al.}(2004)Zhang, Snyder and Sun}]{zhang_2004}
Zhang F, Snyder C, Sun J. 2004. Impacts of initial estimate and observation
  availability on convective-scale data assimilation with an ensemble {Kalman}
  filter. \emph{Monthly Weather Review} \textbf{132}: 1238--1253,
  \doi{10.1175/1520-0493(2004)132<1238:IOIEAO>2.0.CO;2}.

\end{thebibliography}

\appendix
%\appendixtitle{Detailed derivations for the ETKPF}

%%%%%%%%%%%%%%%%%%%%%%%%%%%%%%%%%%%%%%%%%%%%%%%%%%%%%%%%%%%%%%%%%%%%%
\section{Riccati equation for the transform filter}
\label{ap:riccati}
First let us write \cref{eq:ricc} replacing $W^{\epsilon}$ with $X$ and $(k-1) \tilde P^{a,\gamma}$ with $C$ for more clarity:
\begin{align}
A X + X A' +  XX' - C = 0, \label{eqapp:ricc}
\end{align}
where we transposed the first $X$, which we can do as we seek a symmetric solution. 
Using Newton's method to solve this equation we find the candidate $X_{n+1}$ recursively by solving 
\begin{align}
(A + X_n) X_{n+1} + X_{n+1}(A' + X_n) = X_n X_n' + C. \label{eqapp:lyap}
\end{align}
Theorems 9.1.1 and 9.1.2  in  \citet{lancaster_algebraic_1995} show that if the starting value $X_0$ is symmetric and large enough, then \cref{eqapp:lyap} has a unique positive definite solution for all $n$, and the sequence $(X_n)$ converges quadratically to  the largest positive definite
solution of \cref{eqapp:ricc}.

At each step of the algorithm we solve  \cref{eqapp:lyap}
using the $O(k^3)$ algorithm of \citet{bartels_solution_1972}, until a desired level of accuracy is reached, which in our application typically occurs after less than 10 steps.
There are other algorithms besides Newton's method which are more efficient when a high degree of accuracy is desired, but for the present case we are satisfied with this method as it is straightforward to understand and to implement, and it converges in a few steps to a solution accurate enough for our purpose.

To verify that the solution $W^{\epsilon}$ is such that $W^{\epsilon}\bm{1}=0$, first notice that we can pull out a factor $(X^b)'$ from $\tilde L^{\gamma}$ and thus $\bm{1}'\tilde L^{\gamma} = \bm{0}'$ and $\bm{1}' W^{\mu} = \bm{1}'$. $W^{\alpha}$ has only one 1 per column and thus $\bm{1}' W^{\alpha}=\bm{1}'$. Therefore $\bm{1}' A = \bm{0}'$ and $A \bm{1} = \bm{0}$. Because we can pull out a factor $(X^b)'$ on the left and a factor $X^b$ on the right of $\tilde P^a$ we can also see that $\bm{1}' \tilde P^{a,\gamma} \bm{1} = 0$. 
Multiplying \cref{eqapp:ricc} by $\bm{1}'$ from the left and by $\bm{1}$ from the right, it follows that $X'\bm{1}=\bm{0}$ and by symmetry, also $X\bm{1}=0$.

%%%%%%%%%%%%%%%%%%%%%%%%%%%%%%%%%%%%%%%%%%%%%%%%%%%%%%%%%%%%%%%%%%%%%
\section{Efficient computation of weight matrices}
\label{ap:deriv}

The derivation of the algorithm in ensemble space starts by applying Woodbury's formula to compute the inverse in the Kalman gain $\tilde K(\gamma P^b)$ and results in the following expression after some further simplifications:
$$
\tilde K(\gamma P^b) = \gamma \Big((k-1)I + \gamma S\Big)^{-1} 
(H X^b)' R^{-1}.
$$
Using the definition of $Q$ we can then write
$$
\tilde Q = \gamma S \Big((k-1)I + \gamma S\Big)^{-2},
$$
which is correct because the matrices on the right commute. To compute $\tilde K((1-\gamma) Q)$ we substitute the expression for $\tilde Q$ in the definition and apply again Woodbury's formula. After some further simplifications we can find that:
$$
\tilde K((1-\gamma) Q) = (1-\gamma) \Big( I + (1-\gamma) \tilde Q S \Big)^{-1} 
\tilde Q (H X^b)' R^{-1}.
$$

Splitting the ensemble into mean and deviations one can rewrite the $W^{\mu}$ matrix in \cref{eq:Wmu} as
$$
W^{\mu} = I - \tilde L^{\gamma} H X^b + \tilde L^{\gamma} (y - H \bar x^b) \bm{1}',
$$
where the first part will be computed using the $S$ matrix and the last part using the $S$ matrix and the $c$ vector.
Using the expressions for $\tilde K(\gamma P^b)$ and $\tilde K((1-\gamma) Q)$ to compute $\tilde L^{\gamma}$ and some further simplifications, we can derive the first part as
$$
I - \tilde L^{\gamma} H X^b = \Big( I + (1-\gamma) \tilde Q S \Big)^{-1} 
(k-1) \Big((k-1)I + \gamma S\Big)^{-1}.
$$
Finally, using the spectral decomposition of $S$ and basic rules of algebra we can find the rational function 
\begin{align}
f^{\mu}(\lambda) &= \frac{(k-1) \gamma \lambda + (k-1)^2} 
{\gamma\lambda^2 + 2 (k-1) \gamma \lambda + (k-1)^2}. 
\label{eq:fmu}
\end{align}

The second part of the matrix $W^{\mu}$ can be derived similarly as:
\begin{align*}
\tilde L^{\gamma} &(y - H \bar x^b) = \\ &\Big( I + (1-\gamma) \tilde Q S \Big)^{-1} 
\Big( 
(1-\gamma) \tilde Q + \gamma \Big((k-1)I + \gamma S\Big)^{-1} 
\Big) c,
\end{align*}
from which we can find the function $f^{\bar \mu}$ after plugging in the spectral decomposition of $S$:
\begin{align}
f^{\bar \mu}(\lambda)     &= \frac{1} { (k-1) + \gamma \lambda} \cdot
\Bigg( \gamma + 
\frac{ (k-1)  \gamma (1-\gamma) \lambda} 
{\gamma \lambda^2 + 2 (k-1) \gamma \lambda + (k-1)^2}
\Bigg).
\label{eq:fmubar}
\end{align}

Using the expression for $\tilde K((1-\gamma) Q)$ and $\tilde Q$ we can similarly find that 
$$
\tilde P^{a,\gamma} = \Big( I + (1-\gamma) \tilde Q S \Big)^{-1} 
\gamma S \Big((k-1)I + \gamma S\Big)^{-2},
$$
from which $f^{\gamma}$ can easily be found as
\begin{align}
f^{\gamma} (\lambda) = \frac{\gamma \lambda}
{\gamma \lambda^2 + 2 (k-1) \gamma \lambda + (k-1)^2}.
\label{eq:fgam}
\end{align}

For the weights $\alpha^{\gamma,i}$ the derivation is similar and we find that they are proportional to
\begin{align*}
\exp \Bigg( -\frac{1}{2} \Big(
&(k-1)^2 (1-\gamma) \Big((k-1)I + \gamma S\Big)^{-2} \cdot \\
&\Big( I + (1-\gamma) \tilde Q S \Big)^{-1}
(S - c \bm{1}')
\Big)_{ii} \Bigg).
\end{align*}
The final expression can be found by developing the last product in the exponential and by substituting the spectral decomposition of $S$, which results in the following:
\begin{align}
f^{\alpha} (\lambda) =
\frac{ (k-1)^2 (1-\gamma)}
{\gamma \lambda^2 + 2 (k-1) \gamma \lambda + (k-1)^2}.
\label{eq:falpha}
\end{align}

One can easily see 
what happens in the limiting cases of $\gamma=0$ and $\gamma=1$. 
Setting $\gamma=0$ gives $f^{\mu}(\lambda)= f^{\alpha}(\lambda)=1$
and $f^{\bar \mu}(\lambda)=f^{\gamma}(\lambda)=0$. Hence $W^\mu=U U' = I$,
$W^\epsilon=0$ and 
$\alpha_i \propto \exp(-\frac{1}{2}U_{ii} + c_i) \propto \ell(x_i|y)$.
Hence the resulting analysis is equivalent to the PF.

In the case where $\gamma=1$, $f^{\alpha}(\lambda)=0$ and thus 
$\alpha^{\gamma,i}=\frac{1}{k}$ and $W^{\alpha}=I$. Furthermore,
$f^{\mu}(\lambda)$ simplifies to $(k-1)/((k-1) + \lambda),$ 
and $f^{\gamma}(\lambda)$ to $\lambda/( (k-1) + \lambda)^2$.
The CARE in \cref{eq:ricc} has thus the positive semidefinite
solution $W^\epsilon= U\delta(f^\epsilon(\bm{\lambda}))U'$ where
%$$
%2 f^\epsilon(\lambda) f^\mu(\lambda) + f^\epsilon(\lambda)^2 = f^\gamma(\lambda)
%$$
$$
2 f^\epsilon(\lambda) f^\mu(\lambda) + f^\epsilon(\lambda)^2 = (k-1) f^\gamma(\lambda)
$$
or
%\begin{align}
%f^\epsilon(\lambda)= \frac{\sqrt{(k-1)^2 + \lambda^2} - (k-1)}
%{\lambda+k-1}.
%\end{align}
\begin{align}
f^\epsilon(\lambda)= \frac{\sqrt{(k-1)( (k-1) + \lambda)} - (k-1)}
{ ( (k-1) + \lambda)}
\end{align}

The sum $W^\mu + W^\epsilon$ is thus given by
%$$U \delta\Big( \sqrt{\frac{\sqrt{k-1}}{\sqrt{(k-1) + \bm{\lambda}} } } \Big)U',$$ 
$$U \delta\Big( \sqrt{\frac{(k-1)}{(k-1) + \bm{\lambda} } } \Big)U',$$ 
which is the formula for the transformation matrix in the ETKF. 
%The sum $W^\mu + W^\epsilon$ is thus given by
%$U \delta( (\bm{\lambda} + k -1)^{-1/2}) U'$ which agrees with the well-known
%formula for the transformation matrix in the ETKF.

\end{document}